\documentclass[useAMS,usenatbib]{mn2e}
\usepackage{graphicx, setspace, subfigure, latexsym, amssymb, amsmath, booktabs, wasysym, paralist, longtable}
\usepackage{multirow}
\usepackage{gensymb}

\title[HTRU VI: An ANN and 75 Normal Pulsars]{The High Time Resolution Universe Survey VI: An Artificial Neural Network and Timing of 75 Pulsars}
\author[S. D. Bates et al.]
	{S. D. Bates$^{1,2}$, M. Bailes$^{3,4}$, B. R. Barsdell$^{3,4}$, N. D. R. Bhat$^{3,4}$, M. Burgay$^{5}$,\newauthor S. Burke-Spolaor$^{6}$, D. J. Champion$^7$, P. Coster$^{3}$, N. D'Amico$^{5}$, A. Jameson$^{3,4}$,\newauthor S. Johnston$^{8}$, M. J. Keith$^{8}$, M. Kramer$^{7,2}$, L. Levin$^{3,4,8}$, A. Lyne$^{2}$, S. Milia$^{5,9}$, C. Ng$^{7}$,\newauthor C. Nietner$^{1}$, A. Possenti$^{5}$, B. Stappers$^{1}$, D. Thornton$^{1}$ and W. van Straten$^{3,4}$\\
$^{1}$Jodrell Bank Centre for Astrophysics, School of Physics and Astronomy, The University of Manchester, Manchester M13 9PL, UK\\
$^{2}$Department of Physics, West Virginia University, Morgantown, WV 26506, USA\\
$^{3}$Centre for Astrophysics and Supercomputing, Swinburne University of Technology, PO Box 218 Hawthorn, VIC 3122, Australia\\
$^{4}$The ARC Centre of Excellence for All-Sky Astrophysics (CAASTRO)\\
$^{5}$INAF - Osservatorio Astronomico di Cagliari, Poggio dei Pini, 09012 Caopterra, Italy\\
$^{6}$NASA Jet Propulsion Laboratory, M/S 138-307, Pasadena CA 91106, USA\\
$^{7}$MPI fuer Radioastronomie, Auf dem Huegel 69, 53121 Bonn, Germany\\
$^{8}$Australia Telescope National Facility, CSIRO, P.O. Box 76, Epping NSW 1710, Australia\\
$^{9}$Dipartimento di Fisica, Universit\`{a} degli Studi di Cagliari, Cittadella Universitaria, 09042 Monserrato (CA), Italy}
\begin{document}

\date{Accepted; \today}

\pagerange{\pageref{firstpage}--\pageref{lastpage}} \pubyear{2012}

\maketitle

\label{firstpage}
\begin{abstract}
We present 75 pulsars discovered in the mid-latitude portion of the High Time Resolution Universe survey, 54 of which have full timing solutions. All the pulsars have spin periods greater than 100~ms, and none of those with timing solutions are in binaries. Two display particularly interesting behaviour; PSR~J1054--5944 is found to be an intermittent pulsar, and PSR~J1809--0119 has glitched twice since its discovery.

In the second half of the paper we discuss the development and application of an artificial neural network in the data-processing pipeline for the survey. We discuss the tests that were used to generate scores and find that our neural network was able to reject over 99\% of the candidates produced in the data processing, and able to blindly detect 85\% of pulsars. We suggest that improvements to the accuracy should be possible if further care is taken when training an artificial neural network; for example ensuring that a representative sample of the pulsar population is used during the training process, or the use of different artificial neural networks for the detection of different types of pulsars.
\end{abstract}


\section{Introduction}
\subsection{The High Time Resolution Universe survey}
While the known pulsar population now stands at over 2000 pulsars, there are continuing efforts to discover yet more of these fascinating objects. The focus of recent surveys is often on the discovery of millisecond pulsars (MSPs) to be used in pulsar timing arrays for the detection of gravitational radiation \citep{hobbs2009,ferdman2010,jenet2009}, or for more exotic flavours of neutron stars such as rotating radio transients \citep[RRATS,][]{rrats2006} which are not as well studied as the currently known pulsar population. However, the long-anticipated discovery of a binary system containing both a pulsar and a black hole, which would enable high-precision tests of General Relativity \citep{kramer2004}, is unlikely to contain such an exotic pulsar. Instead, the system is likely to contain an ordinary pulsar with period $\sim 1\,\rm{s}$ \citep{fgl2010}. Therefore, the discovery of normal pulsars, with pulse periods greater than 100~ms and period derivatives between $10^{-17}$ and $10^{-13}$, continues to be of great importance. There is also the potential for discovery of new pulsar sub-classes, with behaviour different from those which have come before, e.g.\,the discovery of an intermittent pulsar by \citet{kramer2006a}.

The population of normal pulsars provides a large sample from which meaningful statistics can be drawn \citep{lorimer2011}. These statistics can then be applied in numerous ways, for example;
\begin{itemize}
 \item To provide data against which models of the evolution of pulsars can be tested \citep[e.g.][]{fk06}
 \item As indicators of other astrophysical phenomena, for example the rate of supernova explosions required to produce the observed population \citep[e.g.][]{ridley2010}, or the birthrate of neutron stars in the Galaxy \citep{keane2008}
 \item As probes of the inter-stellar medium. Radio pulses are dispersed as they travel along the line of sight to Earth, and this can be used to `map' the distribution of free electrons along different lines of sight in the Galaxy \citep[though only if there is an independent measure of the pulsar's distance, e.g.][]{lmt85,taylor1993,ne2001}
 \end{itemize}
 
Studies of the properties of the pulsar population also provide insight into the physical processes occurring in the magnetosphere of the pulsar, from which the radio emission originates, and inside the crust of the neutron star. Due to the large diversity in the pulsar population, individual pulsars can sometimes place new constraints on the emission processes, for example, the first known intermittent pulsar --- mentioned earlier --- PSR~B1931+24 \citep{kramer2006a}, is not only observed to switch between observable and non-observable states, but the spin-down rate of the pulsar is observed to increase when the pulsar is emitting. This provided insight into the plasma currents and charge densities inside the pulsar magnetosphere.

Long-term radio timing by \citet{lyne2010} has recently demonstrated that the phenomena of nulling, mode changing and timing noise are related and, likely, due to changes in the pulsar's magnetosphere. Glitches, which conversely occur on very short timescales, are observed as sudden jumps in the rotational frequency of pulsars, and are thought to be caused by a transfer of angular momentum from the interior of the neutron star to its crust. Glitches are most commonly observed to occur in those pulsars with characteristic ages $\tau_\mathrm{c}\sim 10~\rm{kyr}$ \citep{espinoza2011}.

With the numerous applications of a large population of known pulsars, and the issues that remain with models of pulsar emission and neutron star interiors, the discovery of normal pulsars adds strength to the case for further pulsar surveys with current and future radio telescopes. 

The High Time Resolution Universe survey \citep{keith2010} using the Parkes 64-metre radio telescope has, heretofore, resulted in the discovery of both normal pulsars and MSPs \citep[see][]{batesmsps, bailesplanet, htru3, keith2011} and is expected to continue to do so as more data are processed. However, the discovery of normal pulsars, with pulse periods greater than 100~ms, has also continued due to the improved time and frequency resolution, and hence lower sensitivity thresholds, offered by modern hardware.

\begin{table}
	\begin{center}
	\caption{Observational parameters for the mid-latitude portion of the HTRU survey.}
		\begin{tabular}{lr}
		\toprule
		%\multicolumn{2}{c}{Receiver Parameters} \\
		Number of beams  & 13 \\
		Polarizations/beam   & 2 \\
		Centre Frequency & 1352 MHz\\
		Frequency channels  & 1024 $\times$ 390.625 kHz$^\star$ \\
		\midrule
		%\multicolumn{2}{c}{Survey Parameters} \\
		Galactic longitude range & $-120\degree$ to $30\degree$ \\
		Galactic latitude range & $|b| \leq 15\degree$ \\
		Sampling interval & 64 $\mathrm{\mu}$s \\
		Bits/sample & 2 \\
		Observation time/pointing & 540 s \\
		\bottomrule
		\end{tabular}
		\label{table:survey}
	\end{center}
	$^\star$154 of these channels are masked to remove interference
\end{table}

\subsection{Candidate Selection in Pulsar Surveys}
Modern pulsar surveys produce vast quantities of data; but once this data has been processed, there are still large numbers of candidate plots which must be inspected by eye to find previously unknown pulsars. For example, the HTRU survey pipeline \citep[see][for details]{keith2010} generates 100 candidates per beam. With over half a million individual observations required to complete the survey, $\sim6 \times 10^7$ candidates could easily be produced by the standard analysis of the data. These candidates are usually inspected by eye, which can be a slow process and also introduces the possibility of human error. 

To make this task manageable, it has always been common to reduce the number of candidates by setting thresholds in signal-to-noise ratio, or by using graphical plotting programs such as \textsc{JReaper} \citep{kel+09}. These programs can be used either to identify regions of parameter space where good-quality candidates are likely to be found, or where candidates are not likely to be genuine, for example due to radio-frequency interference (RFI). The problem with such techniques is that, while they offer relief from the large number of candidates, they make the assumptions that
\begin{inparaenum}[\itshape a\upshape)]
\item a candidate can be rejected based purely on a low S/N; and
\item a candidate can be rejected if it has a period which is related to a known RFI source.
\end{inparaenum}
While these assumptions are not baseless, they also cannot be shown to apply to every candidate in an entire survey, nor do they make use of all the information that is available for each candidate. Indeed, these cuts will often only produce a limited reduction in the number of candidates, while the levels at which cuts are made can vary (e.g.\,due to particularly strong RFI during an observation), making it difficult to be consistent.

\begin{table}
	\begin{center}
	\caption{Observing system details for the timing observations made as part of this work. Note the specifications for the Lovell Telescope take into account the standard removal of a section of the observing bandwidth.}
		\begin{tabular}{lcccc}
		\toprule
		Telescope  & Centre Freq. & BW & N$_\mathrm{chans}$ & $\langle t_\mathrm{obs} \rangle$ \\
		&(MHz) &(MHz) & &(s)\\
		\midrule
		Parkes 64-metre & 1369 & 256 & 1024 & 600 \\
		& & & & \\
		Lovell Telescope & 1524 & 384 & 768 & 900 \\
		\bottomrule
		\end{tabular}
		\label{table:timing}
	\end{center}
\end{table}

The Pulsar Search Collaboratory \citep{rosen2010} tackle this problem by storing candidates from the GBT 350-MHz survey \citep{boyles2010} in an online database, where users can view and rank candidate plots. By distributing the workload, some of the human error is mitigated, however, a large number of people need to be trained to view the candidates, and there will be a lack of consistency between users of the system.

It seems that once future, large-scale, pulsar surveys such as those with the LOw Frequency ARray \citep[LOFAR,][]{bws, stappers2011} and the Square Kilometre Array \citep[SKA,][]{smits2008} begin to produce results, it would be ideal to have the use of automated computer algorithms to identify the best pulsar candidates. 

In particular, computer learning algorithms such as Artificial Neural Networks (ANNs), which are adept at solving problems involving pattern recognition, show promise of providing a way to analyse candidates without the need for human inspection. \citet{kel+09} created several scores to describe pulsar candidates, which resulted in the discovery of a number of low S/N pulsars in the Parkes Multibeam Pulsar Survey \citep[PMPS,][]{mlc+01}. This work, which was continued by \citet{eatough2010} who implemented an ANN during further reprocessing of the PMPS, resulting in the discovery of PSR~J1926+0739 \citep{eatoughthesis}. 

In this paper we present previously unpublished results from the HTRU pulsar survey, outlining the parameters of 75 newly-discovered pulsars, with complete timing solutions for 54. We will then briefly outline the theory behind computer learning algorithms, and discuss the ANN which was trained using early HTRU data and then used as a tool during the data processing.

\begin{figure}
	\begin{center}
	\includegraphics[width=8.5cm]{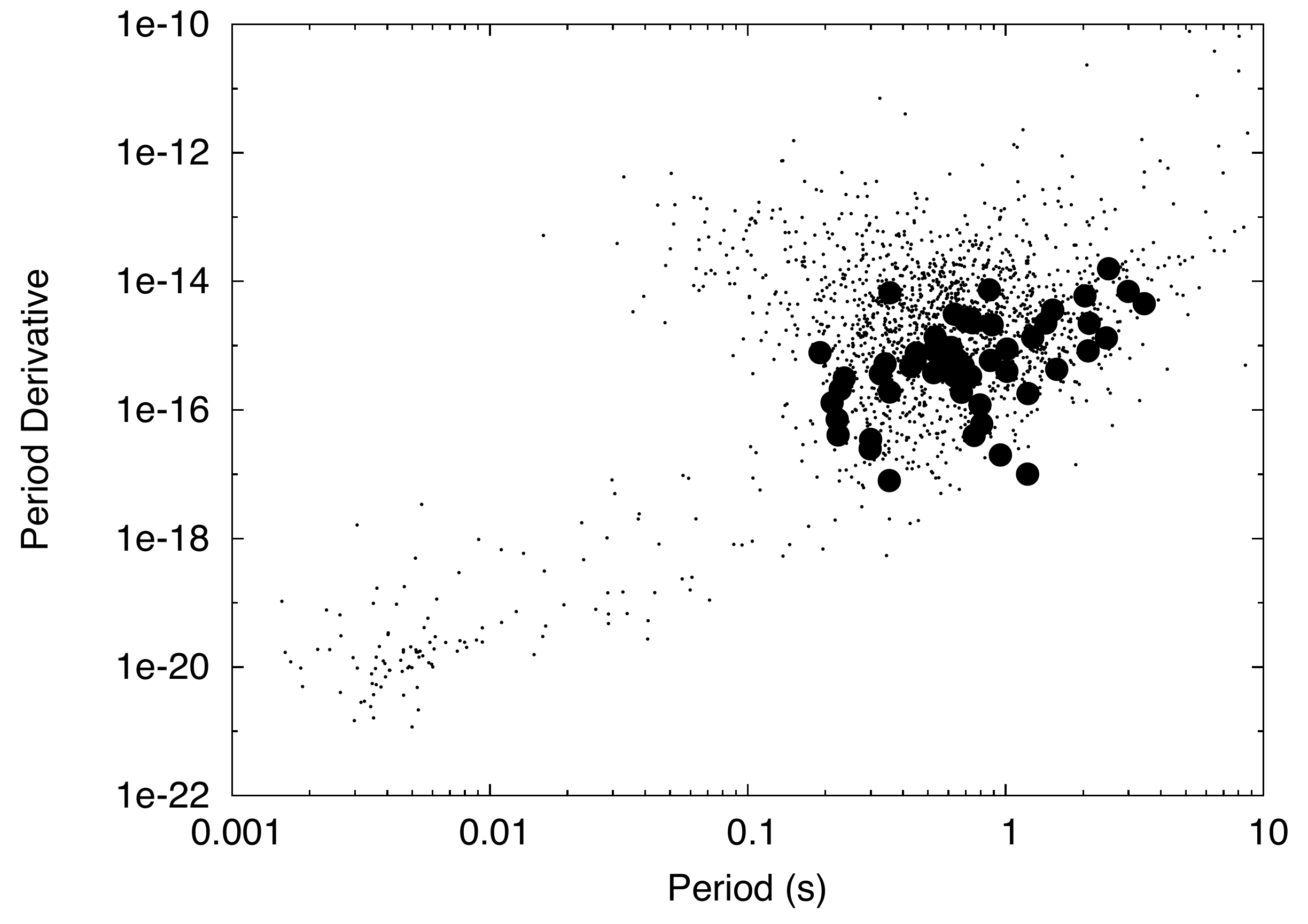}
	\end{center}
	\caption{$P$--$\dot{P}$ diagram of the known pulsar population. The newly-discovered pulsars presented here are indicated by large points.}
	\label{fig:ppdot}
\end{figure}

\begin{figure*}
	\begin{center}
	\includegraphics[width=17cm]{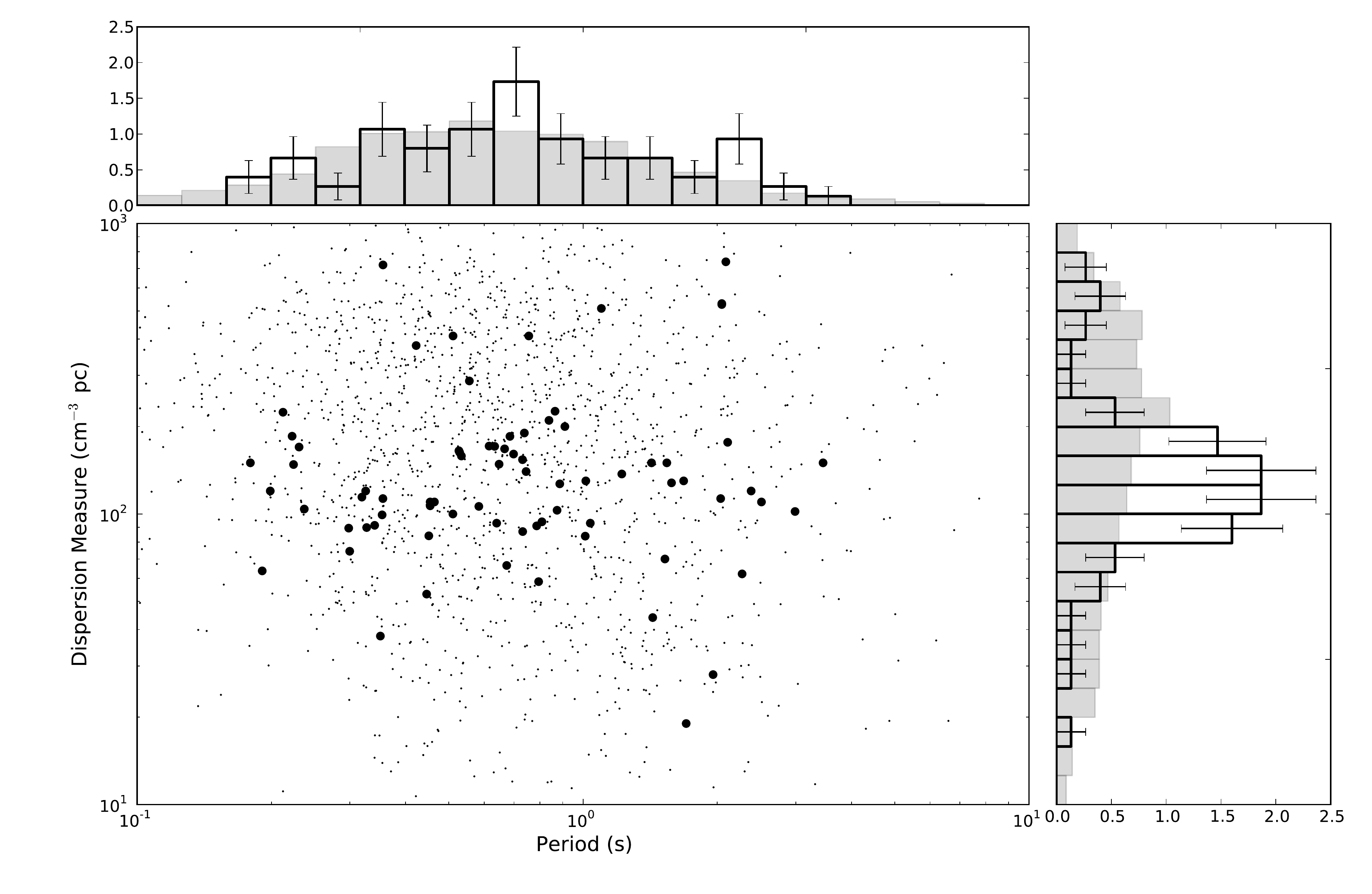}
	\end{center}
	\caption{Period and dispersion measure for all pulsars with periods greater than $0.1~\mathrm{s}$. Small points are previously-known pulsars, taken from the ATNF pulsar catalogue, and large dots are for the new discoveries published here. Also shown, for comparison, are normalised histograms of the $P$ and DM values, in grey for the previously-known pulsars and solid lines for the pulsars published here. Error bars are scaled as $\sqrt{n}$ where $n$ is the number of pulsars in each bin.}
	\label{fig:pdm}
\end{figure*}

\section{Timing results for 75 pulsars in the HTRU survey}
\subsection{Discovery and Timing}
All the pulsars presented here were discovered in the HTRU mid-latitude survey, which has now been fully processed. The survey observed the Galactic plane in the region $-120\degree < l <30\degree$ and $b \leq 15\degree$. A short summary of the survey parameters is given in Table~\ref{table:survey}; see \citet{keith2010} for more details. After the discovery and subsequent confirmation observations with the Parkes 64-metre radio telescope, pulsars with declinations $\delta > -35\degree$ were regularly observed using the 76-metre Lovell Telescope and those below this declination were observed as part of the HTRU timing program at Parkes. 

Timing observations were made using digital filterbanks (DFBs), and were performed approximately once every 3 weeks at Jodrell Bank Observatory (JBO), and once per month at Parkes, using the system parameters outlined in Table~\ref{table:timing}. Timing solutions were obtained using the \textsc{Tempo2} pulsar timing package \citep{tempo2_1}, and are shown in Table~\ref{table:fullsolns} for those pulsars with timing data spanning over 300 days. Parameters which may be derived from these solutions are given in Table~\ref{table:derived}. Those pulsars with a shorter data span, for which we do not yet have a full timing solution, are presented in Table~\ref{table:partsolns} with interim names and only basic parameters.

\begin{table*}
\caption{Observable parameters for each of the pulsars with a full timing solution. Errors in position, period, period derivative and dispersion measure are the 1-sigma errors as reported by \textsc{Tempo2}.}
\begin{tabular}{llllrrr}
\toprule
\multicolumn{1}{c}{Pulsar} & \multicolumn{1}{c}{RA} & \multicolumn{1}{c}{Dec} & \multicolumn{1}{c}{$P$} & \multicolumn{1}{c}{$P$ Epoch} & \multicolumn{1}{c}{$\dot{P}$} & \multicolumn{1}{c}{DM} \\
& \multicolumn{1}{c}{(J2000)} & \multicolumn{1}{c}{(J2000)} & \multicolumn{1}{c}{(s)} & \multicolumn{1}{c}{(MJD)} & \multicolumn{1}{c}{($\times 10^{-15}$)} & \multicolumn{1}{c}{($\rm{cm}^{-3}\,\rm{pc}$)} \\
\midrule
J0807$-$5421  &  08:07:47.185(8)  &  $-$54:21:26.46(8)  &  0.52664353143(3)  &  55333  &  0.378(1)  &  165.03(7) \\
J0905$-$6019  &  09:05:15.245(5)  &  $-$60:19:22.06(3)  &  0.340854176542(8)  &  55191  &  0.5220(3)  &  91.4(4) \\
J0912$-$3851  &  09:12:42.70(2)  & $-$38:51:03(1)  &  1.526085076(3)  &  55093  &  3.59(5) &  70(1) \\
J0919$-$6040  &  09:19:27.87(7)  &   $-$60:40:50.4(3)   &   1.2169757230(6)  &  55190  &  0.01(2)   &  82.5(3)  \\
J0949$-$6902  &  09:49:20.567(6)  &  $-$69:02:41.60(3)  &  0.64001572416(1)  &  55195  &  0.6370(5)  &  93.0(1) \\
J1036$-$6559  &  10:36:20.04(2)  &  $-$65:59:09.27(6)  &  0.53350188629(6)  &  55010  &  1.362(1)  &  158.36(9) \\
J1054$-$5946  & 10:54:30.46(1) & $-$59:46:31.0(1) & 0.228324249982(8) & 55337 & 0.2090(3) & 253.9(6) \\
J1143$-$5536  &  11:43:09.79(2)  &  $-$55:36:04.5(1)  &  0.68535848563(4)  &  55213  &  0.485(2)  &  185.0(1) \\
J1237$-$6725  &  12:37:26.0(2)  &  $-$67:25:34.6(6)  &  2.110974776(2)  &  55185  &  2.23(7)  &  176.5(3) \\
J1251$-$7407  &  12:51:52.94(1)  &  $-$74:07:15.04(9)  &  0.32705773823(2)  &  55332  &  0.3651(8)  &  89.81(5) \\
\\
J1331$-$5245  &  13:31:00.01(4)  &  $-$52:45:25.4(5)  &  0.6481166471(2)  &  55195  &  0.510(9)  &  148.4(3) \\
J1346$-$4918  &  13:46:22.35(2)  &  $-$49:18:07.2(1)  &  0.2996251068(2)  &  55000  &  0.035(3)  &  74.42(7) \\
J1409$-$6953  &  14:09:16.9(1)  &  $-$69:53:34.4(5)  &  0.5285907792(3)  &  55191  &  0.84(1)  &  163(2) \\
J1416$-$5033  &  14:16:44.6(2)  &  $-$50:33:17(3)  &  0.794882546(2)  &  55337  &  0.12(5)  &  58.5(3) \\
J1432$-$5032  &  14:32:52.27(7)  &  $-$50:32:17.3(6)  &  2.0349894792(3)  &  54842  &  5.924(8)  &  113(1) \\
J1443$-$5122  &  14:43:26.97(6)  &  $-$51:22:26(1)  &  0.7320612647(5)  &  54800  &  0.338(9)  &  87.0(7) \\
J1517$-$4636  &  15:17:29.376(9)  &  $-$46:36:00.6(2)  &  0.88661249686(5)  &  55210  &  2.098(2)  &  127.0(1) \\
J1534$-$4428  &  15:34:52.00(5)  &  $-$44:28:09.4(8)  &  1.2214259588(3)  &  55337  &  0.18(2)  &  137.3(2) \\
%J1539$-$4828  &  15:39:40.89(5)  &  $-$48:28:58(2)  &  1.2728416429(9)  &  55071  &  1.34(5)  &  118(2)\\
J1551$-$4424  &  15:51:48.02(5)  &  $-$44:24:42(1)  &  0.6740603610(2)  &  55225  &  0.188(8)  &  66.5(4) \\
J1607$-$6449  &  16:07:48.711(8)  &  $-$64:49:43.08(8)  &  0.298116357616(9)  &  55192  &  0.0249(3)  &  89.39(7) \\
\\
J1612$-$5805  &  16:12:27.816(7)  &  $-$58:05:29.2(1)  &  0.61552045802(3)  &  54893  &  0.9347(9)  &  171.3(4) \\
J1622$-$3751  &  16:22:04.58(4)  &  $-$37:51:13.9(9)  &  0.7314627228(5)  &  55070  &  2.57(1)  &  153.8(5) \\
J1625$-$4913  &  16:25:16.41(2)  &  $-$49:13:44.6(4)  &  0.35585626277(5)  &  54895  &  6.647(1)  &  720(1) \\
J1626$-$6621  &  16:26:06.851(9)  &  $-$66:21:15.27(8)  &  0.45086776633(1)  &  55195  &  0.7664(5)  &  84.11(5) \\
J1627$-$5936  &  16:27:52.59(4)  &  $-$59:36:55.3(2)  &  0.35423394051(6)  &  55188  &  0.008(3)  &  99.3(2) \\
J1629$-$3636  &  16:29:35.81(9)  &  $-$36:36:13(2)  &  2.988192686(9)  &  55000  &  7.0(1)  &  101(1) \\
J1634$-$5640  &  16:34:19.17(2)  &  $-$56:40:48.7(3)  &  0.22420119106(8)  &  55010  &  0.041(2)  &  148.0(1) \\
J1647$-$3607  &  16:47:46.51(2)  &  $-$36:07:04(1)  &  0.21231640921(5)  &  54984  &  0.129(2)  &  224(1) \\
J1648$-$6044  &  16:48:51.23(2)  &  $-$60:44:25.5(1)  &  0.58376499689(5)  &  55222  &  0.429(3)  &  106.2(1) \\
J1700$-$4422  &  17:00:53.67(8)  &  $-$44:22:27(1)  &  0.7555354095(3)  &  55065  &  0.04(2)  &  410(9) \\
\\
J1705$-$4331  &  17:05:35.914(7)  &  $-$43:31:13.6(1)  &  0.22256110261(2)  &  54986  &  0.0712(5)  &  185.24(5) \\
J1705$-$6135  &  17:05:15.3(2)  &  $-$61:35:15(2)  &  0.808546089(1)  &  54896  &  0.06(4)  &  94(7) \\
J1709$-$4401  &  17:09:41.39(3)  &  $-$44:01:11.2(6)  &  0.8652353343(7)  &  55000  &  7.37(1)  &  225.8(4) \\
J1710$-$2616 & 17:10:04.9(1) & $-$26:16:35(20) & 0.954158007(1) & 55070 & 0.02(2) & 111(1) \\
J1716$-$4711  &  17:16:01.109(7)  &  $-$47:11:00.9(3)  &  0.55582421598(6)  &  55185  &  0.833(2)  &  287.06(6) \\
%J1719-2334  &  17:19:36.8(6)  &  -23:28:52(1011)  &  0.4539925066(5)  &  55770  &  2.9(7)e$-$15  &  106.7(9) \\
J1720$-$2446  &  17:20:22.46(6)  &  $-$24:46:27(12)  &  0.87426457245(8)  &  55326  &  0.593(4)  &  103(3) \\
%J1729$-$2117  &  17:29:10.816(4) & $-$21:17:26.4(9)  & 0.066292899264(1) & 55505 & 0.00027(6) & 34.24(2) \\
J1733$-$5515  &  17:33:00.4(3)  &  $-$55:15:40(5)  &  1.011233535(8)  &  55194  &  0.4(2)  &  83.9(8) \\
J1744$-$5337  &  17:44:38.92(4)  &  $-$53:37:51(2)  &  0.3556658488(8)  &  55000  &  0.19(1)  &  113(1) \\
J1745$-$3812  &  17:45:15.42(4)  &  $-$38:12:07.3(9)  &  0.6983528638(2)  &  55330  &  2.426(7)  &  160.8(4) \\
J1747$-$1030  &  17:47:58.31(6)  &  $-$10:30:05(4)  &  1.5787928888(2)  &  55509  &  0.43(2)  &  128(7) \\
\\
J1749$-$4931  &  17:49:23.77(4)  &  $-$49:31:59(2)  &  0.445822307(2)  &  55000  &  0.59(2)  &  53(2) \\
J1754$-$2422  &  17:54:36.56(6)  &  $-$24:22:24(49)  &  2.0902480768(4)  &  55310  &  0.83(2)  &  738(6) \\
J1755$-$0903  &  17:55:10.364(5)  &  $-$09:03:51.6(2)  &  0.190709642575(4)  &  55536  &  0.7809(3)  &  63.7(2) \\
J1759$-$1029  &  17:59:34.30(4)  &  $-$10:29:57(3)  &  2.5122628118(5)  &  55348  &  15.74(2)  &  110(10) \\
%J1802-0521  &  18:02:09(3)  &  -05:23:15(117)  &  1.68057266(4)  &  55738  &  8(8)e$-$15  &  128(9) \\
J1802$-$3346  &  18:02:55.2(1)   &  $-$33:46:45(5)   &  2.461051995(3)   & 54894   &  1.32(9) &  217(5) \\
J1803$-$3329  &  18:03:44.453(4)  &  $-$33:29:10.7(3)  &  0.633411983159(4)  &  55152  &  0.3372(2)  &  170.9(6) \\
J1805$-$2948   & 18:05:42.49(1)  &  $-$29:48:00(2)  &  0.4283409894(2)    &  55137  &  0.474(5)  & 167.9(9) \\
J1809$-$0119  &  18:09:51.36(1)  &  $-$01:19:29.0(4)  &  0.7449764016(3)  &  55254  &  2.29(2)  &  140(2) \\
J1811$-$4930  &  18:11:27.19(1)  &  $-$49:30:20.8(2)  &  1.4327041968(1)  &  54996  &  2.254(5)  &  44.0(5) \\
J1812$-$2748  &  18:12:40.58(1)  &  $-$27:48:03(2)  &  0.236983307439(9)  &  55160  &  0.3156(4)  &  104(2) \\
\\
J1812$-$3039 & 18:12:44.902(9) & $-$30:39:21(1) & 0.58747677594(2) & 55336 & 0.6602(8) & 138.9(9) \\
J1814$-$0521  &  18:14:26.13(2)  &  $-$05:21:37.0(8)  &  1.01421948495(6)  &  55257  &  0.884(3)  &  130(2) \\
%J1816-1938  &  18:17:06(1)  &  -19:37:52(160)  &  2.046837651(4)  &  55767  &  1(2)e$-$15  &  526(3) \\
%J1846$-$4244  &  18:46:32.97(4)  &  $-$42:44:36.8(9)  &  2.2733025120(9)  &  55336  &  1.09(3)e$-$15  &  62.2(4) \\ HI LAT!!!
J1854$-$1557  &  18:54:53.6(1)  &  $-$15:57:47(14)  &  3.4531211813(7)  &  55124  &  4.52(4)  &  150(17) \\
%J1901-1739  &  19:01:16(3)  &  -17:37:41(253)  &  1.95685790(3)  &  55699  &  5.0(3)e$-$16  &  28(4) \\
J1907$-$1532  &  19:07:06.78(1)  &  $-$15:32:14.9(8)  &  0.63223532885(4)  &  55424  &  3.084(2)  &  72.6(7) \\
\bottomrule
\end{tabular}
\label{table:fullsolns}
\end{table*}

\begin{table*}
\caption{Derived parameters for each of the pulsars with a full timing solution, based on the values in Table~\ref{table:fullsolns}. Estimates of the distance are based upon a Galactic electron density model by \citet{ne2001}.}
\begin{tabular}{lrrrrrrrrrrrrrr}
\toprule
\multicolumn{1}{c}{Pulsar} & \multicolumn{3}{c}{$l$} & \multicolumn{3}{c}{$b$} & \multicolumn{1}{c}{$d$} & \multicolumn{1}{c}{$\tau_c$} & \multicolumn{3}{c}{$B_\mathrm{surf}$} & \multicolumn{3}{c}{$\dot{E}$} \\
\multicolumn{1}{c}{} & \multicolumn{3}{c}{(deg)} & \multicolumn{3}{c}{(deg)} & \multicolumn{1}{c}{(kpc)} & \multicolumn{1}{c}{(Myr)} & \multicolumn{3}{c}{($10^{11}$ G)} & \multicolumn{3}{c}{($10^{32}$ erg s$^{-1}$)} \\
\midrule
J0807$-$5421 && 268.7 &&& $-$11.6 &&	0.26 &  22 & &  4.5& &&  1.0 &\\
J0905$-$6019 && 278.2 &&& $-$8.8 &&	2.9	 &  10 & &4.2 &  & & 5.2 &\\
J0912$-$3851 && 263.2 &&& 6.6 &&	0.52 &  6.7 & &  23& &&  0.40 &\\
J0919$-$6040 && 279.7 &&& $-$7.8 &&	2.5	  &  1900 & &1.1 &  &&  0.0022 &\\
J0949$-$6902 && 287.8 &&& $-$11.7 && 2.9    &  16 & &  6.4&  & &  0.96 &\\
J1036$-$6559 && 289.8 &&& $-$6.6 && 4.0   &  6.2 &  & 8.5 &  &&  3.5 &\\
J1054$-$5946 && 288.7 &&& $-$0.2 && 4.6   &  17 & &2.2& & & 6.9 &\\
J1143$-$5536 && 293.3 &&& 6.0 && 4.5  &  22 &  & 5.8& &&  0.60 &\\
J1237$-$6725 && 301.6 &&& $-$4.6 && 3.9   &  15 & &  22 &  &&  0.094 &\\
J1251$-$7407 && 303.0 &&& $-$11.2 && 2.4    &  14 & &3.5&  & &  4.1 &\\
&\\
J1331$-$5245 && 309.0 &&& 9.6 && 4.2  &  20 &  & 5.7 &  &&  0.74 &\\
J1346$-$4918 && 312.1 &&& 12.6 && 2.0   &  140 &  & 1.0 &  &&  0.51 &\\
J1409$-$6953 && 309.6 &&& $-$8.0 && 4.3   &  10 &  & 6.7 &  &&  2.2 &\\
J1416$-$5033 && 316.5 &&& 10.1 && 1.5   &  100 & &  3.1 &  &&  0.094 &\\
J1432$-$5032 && 318.9 &&& 9.2 && 2.8  &  5.4 &  & 35 &  &&  0.28 &\\
J1443$-$5122 && 320.1 &&& 7.7 && 1.9  &  34 & &  5.0  & &&  0.34 &\\
J1517$-$4636 && 327.4 &&& 9.2 && 3.2  &  6.7 & &  14  & &&  1.2 &\\
J1534$-$4428 && 331.2 &&& 9.3 && 3.9  &  110 & &  4.7 &  &&  0.039 &\\
%J1539$-$4828 && 329.4 &&& 5.5 && 3.6  &  15 &  & 13 &  &&  0.26 &\\
J1551$-$4424 && 333.6 &&& 7.5 && 2.4  &  57 &  & 3.6 & &&  0.24 &\\
J1607$-$6449 && 322.0 &&& $-$9.5 && 2.1 &  190 &  & 0.86 &  & & 0.37 &\\
\\
J1612$-$5805 && 327.0 &&& $-$5.0 && 3.6 &  10 &  & 7.6 &  & & 1.6 &\\
J1622$-$3751 && 342.3 &&& 8.4 && 3.9  &  4.5 & &  14 &  & & 2.6 &\\
J1625$-$4913 && 334.6 &&& 0.0 && 7.7  &  0.85 & &  15 &  &&  58 &\\
J1626$-$6621 && 322.2 &&& $-$11.9 && 2.2 &  9.3 &  & 5.9 &  &&  3.3 &\\
J1627$-$5936 && 327.3 &&& $-$7.4 && 2.2 &  700 & &  0.53& &&  0.071 &\\
J1629$-$3636 && 344.3 &&& 8.2 && 2.4 &  6.8 &  & 46 &  &&  0.10 &\\
J1634$-$5640 && 330.1 &&& $-$6.1 && 38 &  87 &  & 0.96 & & & 1.4 &\\
J1647$-$3607 && 347.1 &&& 5.8 && 5.2  &  26 &  & 1.7 & &&  5.3 &\\
J1648$-$6044 && 328.2 &&& $-$10.1 && 2.6  &  22 & &  5.0 &  &&  0.85 &\\
J1700$-$4422 && 342.2 &&& $-$1.4 && 5.9 &  300 &  & 1.7 &  &&  0.037 &\\
\\
J1705$-$4331 && 343.4 &&& $-$1.5 && 3.6 &  50 &  & 1.3 &  &&  2.6 &\\
J1705$-$6135 && 328.8 &&& $-$12.2 && 2.5 &  210 & &  2.2 &  &&  0.045 &\\
J1709$-$4401 && 343.5 &&& $-$2.4 && 4.4 &  1.9 & &  25 & & & 4.5 &\\
J1710$-$2616 && 357.9 &&& 8.0 && 2.6  &  760 & &  1.4 &&  &  0.0091 &\\
J1716$-$4711 && 341.5 &&& $-$5.2 && 7.7 &  11 &  & 6.8 & && 1.9 &\\
J1720$-$2446 && 0.4 &&& 7.0 && 2.3  &  23 &  & 7.2 && & 0.35 &\\
%J1729$-$2117 && 4.5 &&& 7.2 && 1.1  &  3900 & &  0.042 && &  0.37 &\\
J1733$-$5515 && 336.2 &&& $-$11.8 && 2.1  &  40 &  & 6.4 & & &  0.15 &\\
J1744$-$5337 && 338.5 &&& $-$12.4 && 3.1 &  30 &  & 2.6  && &  1.7 &\\
J1745$-$3812 && 352.0 &&& $-$4.8 && 3.3 &  4.6 & &  13  & && 2.8 &\\
J1747$-$1030 && 16.2 &&& 9.0 && 3.5 &  58 &  & 8.2 & &&  0.043 &\\
\\
J1749$-$4931 && 342.5 &&& $-$11.1 && 1.4 &  12 &  & 5.1 && &  2.6 &\\
J1754$-$2422 && 4.9 &&& 0.6 && 11  &  40 &  & 13 & & & 0.036 &\\
J1755$-$0903 && 18.3 &&& 8.2 && 1.8 &  3.9 & &  3.9 & & & 44 &\\
J1759$-$1029 && 17.6 &&& 6.5 && 2.7 &  2.5 & &  63&  & & 0.39 &\\
J1802$-$3346 && 357.7 &&& $-$5.6 && 5.4   &  30 &  & 18 &&&  0.035 &\\
J1803$-$3329 && 358.0 &&& $-$5.6 && 4.1   &  30 & &  4.6 & & &  0.52 &\\
J1805$-$2948 && 1.5 &&& $-$4.2 && 3.8   &  14 & &  4.5 &&&  2.4 &\\
J1809$-$0119 && 27.0 &&& 8.6 && 4.3 &  5.2 &  & 13  & && 2.2 &\\
J1811$-$4930 && 344.2 &&& $-$14.3 && 1.3  &  10 &  & 18  && &  0.30 &\\
J1812$-$2748 && 3.9 &&& $-$4.6 && 2.5  &  12 &  & 2.7 & & &  9.4 &\\
\\
J1812$-$3039 && 1.4 &&& $-$6.0 && 3.5  &  14 & &  6.3 & &&  1.3 &\\
J1814$-$0521 && 23.9 &&& 5.7 && 3.4 &  18 & &  9.5  && & 0.33 &\\
J1854$-$1557 && 19.0 &&& $-$7.9 && 4.4  &  12 & &  40 & && 0.043 &\\
J1907$-$1532 && 20.7 &&& $-$10.4 && 2.1  &  3.2 &  & 14 & && 4.8 &\\

\bottomrule
\end{tabular}
\label{table:derived}
\end{table*}

\subsection{Features of the new discoveries}
The positions of these pulsars in the $P$--$\dot{P}$ diagram are shown in Figure~\ref{fig:ppdot}. All of these pulsars lie in the region of the diagram which contains the normal pulsars, with pulse periods greater than 100~ms, and typical period derivatives of $10^{-14}$ - $10^{-17} \mathrm{s/s}$.

For the millisecond pulsars (MSPs) discovered in the HTRU survey, it is clear that the increased time and frequency resolution over previous surveys allows the discovery of more dispersed, and often more distant, sources compared to previous surveys \citep{batesmsps}. To test whether this is the case for the normal pulsars, we can compare the distribution of DM values in the known population \citep[taken from the ATNF pulsar catalogue,][]{mhth05} with that of the discoveries published here.

Plotting the periods and DMs of these two populations in Figure~\ref{fig:pdm}, the DM distribution of the pulsars in the catalogue appears to peak at a higher DM than for the pulsars discovered in HTRU. This is a result of the lower sensitivity limits, at long pulse periods, in previous surveys.

There is also a contribution to this effect from the pulsar distribution in Galactic latitude, which is skewed towards $|b|<5\degree$, and hence higher DMs, by the large number of pulsars which were discovered in the PMPS. To ensure that this apparent difference in DM distribution is not entirely produced by this effect, pulsar DMs were selected at random from the ATNF pulsar catalogue such that the $b$ distribution of the pulsars matched that in our sample. A two-sided KS test was then performed on this synthesised DM distribution and our sample. Repeating this method 1000 times, it was found that the synthetic distribution tends to peak at a slightly higher value of DM, with the probability of the two distributions being the same calculated to be 0.02.

The two period distributions, however, look very similar. This is as expected, given that at long pulse periods, the additional frequency and time resolution that we have over previous surveys are not a factor.

%\subsection{The low-$\dot{P}$ pulsars PSRs~J1729--2117 and J1812--3039}
%In Figure~\ref{fig:ppdot}, PSRs~J1729--2117 and J1812--3039 stand out as having extremely low values of $\dot{P} \sim 10^{-19}$ compared to the other pulsars presented here. PSR~J1729--2117, in particular, is separated significantly; not only in $\dot{P}$, but also with a shorter pulse period of $\sim 66$~ms.

%As PSR~J1729--2117 is isolated, it satisfies the definition of a `disrupted recycled pulsar', DRP, as given by \citet{belczynski2010}; an isolated pulsar, with pulse period $P>20~\mathrm{ms}$ and surface magnetic field $B_\mathrm{surf} < 3 \times 10^{10}~\mathrm{G}$. The DRPs are believed to have once been members of a binary system with a high-mass companion, which disrupted the binary when it underwent a supernova explosion \citep{lorimer2004}.

%Conversely, PSR~J1812--3039 has such a long rotational period that it seems unlikely that it has ever accreted mass from a companion, and its low magnetic field cannot be explained via this evolutionary path. Instead, it seems this pulsar is one of the $\sim0.3\%$ of non-recycled pulsars with $B_\mathrm{surf} < 3 \times 10^{10}~\mathrm{G}$, as estimated by \citeauthor{belczynski2010} based upon the work of \citet{fk06} and \citet{ridley2010}.

\subsection{The intermittent pulsar PSR~J1054--5946}
During the timing campaign to obtain a solution for PSR~J1054--5946, it was noticed that although this pulsar is relatively bright, often no emission was detected in the folded data. Given that the pulsar's DM is 253.9~$\rm{cm}^{-3}\,\rm{pc}$, it seems extremely unlikely that scintillation could be responsible for such behaviour.

In fact, PSR~J1054--5946 displays behaviour similar to PSR~B1931+24 \citep{kramer2006a} and a handful of other pulsars \citep{obrienthesis, camilo2012}, which are known as ``intermittent pulsars''. Although our timing data are too poorly spaced to draw any conclusions about the possibility of periodicities in the switch in behaviour, we note that PSR~J1054--5946 has been observed to switch from a detectable state to a non-detectable state, and back again, within the space of one day.

\subsection{The glitching pulsar PSR~J1809--0119}
Timing analysis of this pulsar (which rotates with a frequency of 1.34~Hz) revealed two glitches separated by $\sim400~\rm{days}$, which are described in Table~\ref{table:glitches}. With characteristic age $\tau_\mathrm{c} \sim 5.2~\rm{Myr}$, PSR~J1809--0119 is in the oldest 10\% of glitching pulsars \citep{espinoza2011}. Very few pulsars have a characteristic age over 10~Myr, whereas pulsars with younger characteristic ages are observed to glitch more freqeuently.

Further monitoring will reveal whether PSR~J1809--0119 is a frequent glitcher or that having two glitches in our data span in unusual. However, the empirical relationship calculated by \citeauthor{espinoza2011} for the average number of glitches per year,
\begin{equation}
N \simeq 6\tau_\mathrm{c}^{-0.5},
\label{eq:glitches}
\end{equation}
which is $\sim0.1$ for PSR~J1809--0119, suggests that such frequent glitching is unlikely, unless this relationship has been distorted by small glitches that have gone undetected in the known population. Unfortunately the limited S/N of timing observations of this pulsar do not allow us to probe other unusual behaviour of the pulsar such as profile variations or moding, as observed by \citet{weltevrede2011} in the case of PSR~J1119--6127.

\begin{table}
	\begin{center}
	\caption{Parameters for the two glitches observed in PSR~J1809--0119.}
		\begin{tabular}{cccc}
		\toprule
		Glitch  & MJD & $\Delta\nu$ & $\Delta\nu/\nu$ \\
		Number & & ($\mu\mathrm{Hz}$) & ($10^{-9}$) \\
		\midrule
		1 & 55406(2) & 0.0023(3) & 1.7(3) \\ 
		2 & 55803(1) & 0.004(1) & 3.0(4) \\
		\bottomrule
		\end{tabular}
		\label{table:glitches}
	\end{center}
\end{table}

The size of the glitches, characterised by $\Delta\nu/\nu$, are relatively small but \citeauthor{espinoza2011} showed that the glitch size distribution is double-peaked, with the first peak at $\log(\Delta\nu/\nu\,[10^{-9}]) \simeq 0.25$. Since the values of $\log(\Delta\nu/\nu\,[10^{-9}])$ for the two glitches are 0.23 and 0.48, they sit at this first peak in the distribution.

\subsection{Pulse Profiles}
\begin{figure*}
	\begin{center}
	\includegraphics[width=17.5cm]{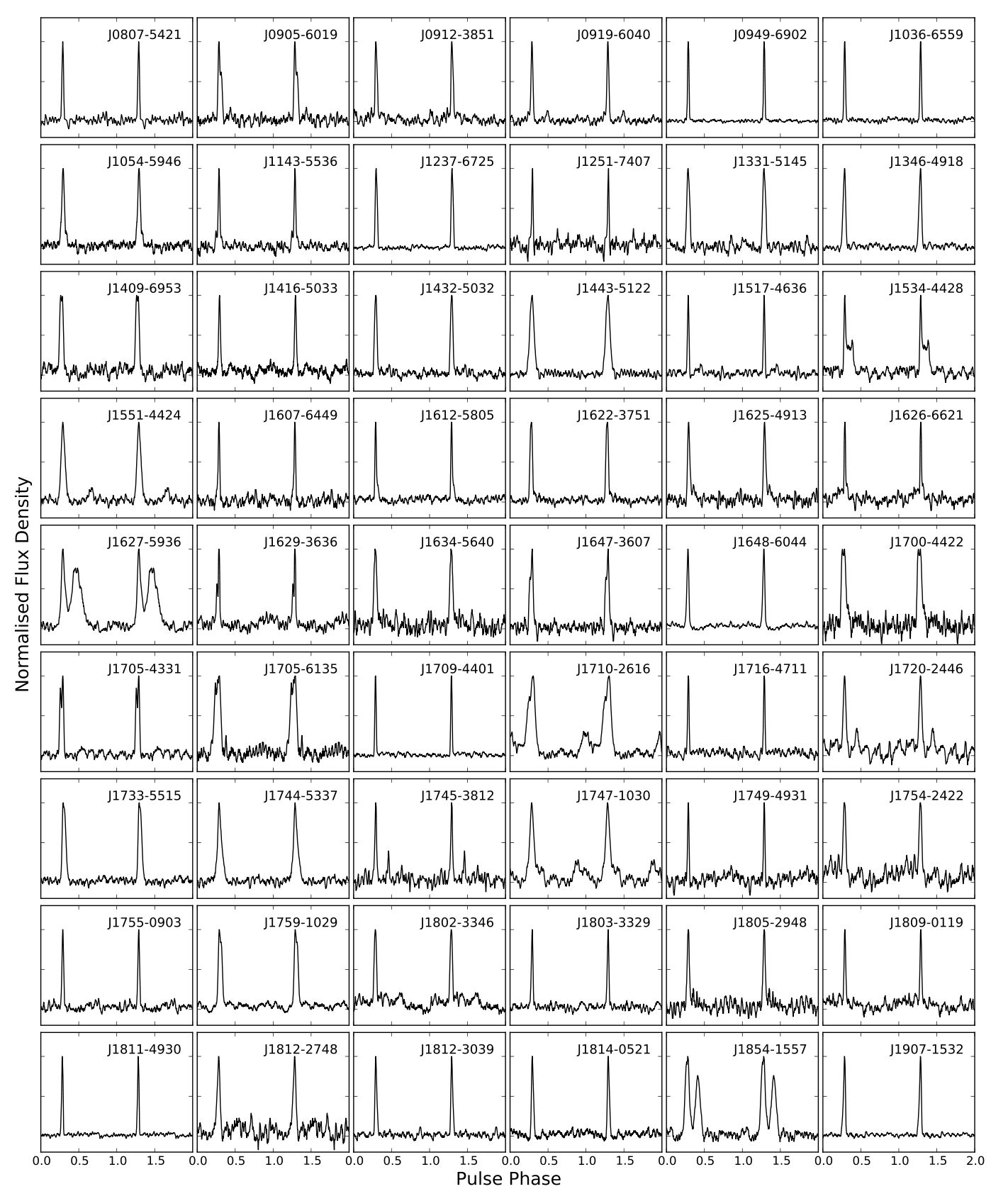}
	\end{center}
	\caption{Pulse profiles at an observing frequency of 1.4~GHz for each of the pulsars with a full timing solution, made by summing several timing observations. Profiles are not flux-calibrated, and the amplitudes have all been normalised to one.}
	\label{fig:profiles}
\end{figure*}

Integrated pulse profiles, obtained from the timing data taken at an observing frequency of 1.4~GHz for each of the 54 pulsars with full timing solutions, are shown in Figure~\ref{fig:profiles}. The data were folded at multiples of the pulse period to ensure that the measured spin frequencies were the fundamental frequencies. For many of the pulsars, the pulse profiles are typical \citep[e.g.][]{lynesmith}, best described by single-peaked pulses with a duty cycle of $\sim$10\%. In some cases (e.g. PSRs J1629--3636 and J1705--4331), the profile is best described by two peaks which have a very small separation, and in others, for example PSRs J1535--4432 and J1627--5933, the two components are very distinct, and form a wide overall pulse shape.

None of the pulsars display evidence of an interpulse trailing the main pulse by $\sim0.5$ in pulse phase. This is not unexpected, since \citet{weltevrede2010} reported that only $\sim2\%$ of the published normal pulsars are observed to have interpulses. 

None of the profiles in Figure~\ref{fig:profiles} displays the classic exponential tail of scattering caused by propagation of the radio signal through the interstellar medium. However, given that \citet{bcc+04} showed that there is significant variation around the relationship between scattering timescale, $\tau$, and DM, we find that our results are in agreement with the predictions of the scattering model.
%None of the profiles in Figure~\ref{fig:profiles} displays the classic exponential tail of scattering caused by propagation of the radio signal through the interstellar medium. Using the Galactic electron density model of \citet{ne2001}, we can predict the scattering timescale, $\tau$, for each of the pulsars. With the exception of PSR~J1625--4913, $\tau$ is predicted to be less than $\sim 1\%$ of the pulse period; however, for PSR~J1625--4913, the predicted value of $\tau$ is $13\%$ of the pulse period. This is in discrepancy with the pulse profile in Figure~\ref{fig:profiles}, where the pulse duty cycle is $5\%$. This pulsar has a particularly high DM of 720~$\rm{cm}^{-3}\,\rm{pc}$, and so there will be significant variation in $\tau$ across the observing bandwidth. In addition, \citet{bcc+04} showed that there is significant variation around the relationship between $\tau$ and DM, and so overall we find that the model agrees with the pulse profiles which have been observed.

\begin{table}
\caption{Basic observable parameters for those pulsars without a fully-determined timing solution. The given RA and Dec reflect the position of the survey pointing in which the pulsar was discovered, not a position from pulsar timing.}
\begin{tabular}{lllrrrr}
\toprule
\multicolumn{1}{c}{Pulsar} & \multicolumn{1}{c}{RA} & \multicolumn{1}{c}{Dec} & \multicolumn{1}{c}{$P$} & \multicolumn{3}{c}{DM} \\
& \multicolumn{1}{c}{(J2000)} & \multicolumn{1}{c}{(J2000)} & \multicolumn{1}{c}{(s)} & \multicolumn{3}{c}{($\rm{cm}^{-3}\,\rm{pc}$)} \\
\midrule
J0835$-$42 & 08:35:37 & $-$42:32:37 & 0.7384 && 190 &\\
J1105$-$43 & 11:05:24 & $-$43:57:01 & 0.3511 && 38&\\
J1132$-$46  & 11:32:33 & $-$46:55:06 &0.3254 && 120&\\
J1530$-$63 & 15:30:52 & $-$63:43:33 & 0.9103 && 200 &\\
J1552$-$62 & 15:52:38 & $-$62:14:31 & 0.1988 && 120 &\\
&\\
J1614$-$38 & 16:14:43 & $-$38:46:15 & 0.4641 && 110 &\\
J1635$-$26 & 16:35:52 & $-$26:16:17 & 0.5105 && 100 &\\
J1638$-$42 & 16:38:31 & $-$42:33:56& 0.5109 && 410 &\\
J1705$-$52 & 17:05:50 & $-$52:36:17 & 0.2307 && 170 &\\
J1719$-$23 & 17:19:37 & $-$23:29:07 & 0.4540 && 110 &\\
&\\
J1757$-$15 & 17:57:24 & $-$15:03:18 & 0.1794 && 150&\\
J1802$-$05 & 18:02:12 & $-$05:23:53 & 1.681 && 130 &\\
J1816$-$19 & 18:16:47 & $-$19:38:30 & 2.047 && 530 &\\
J1818$-$01 & 18:18:15 & $-$01:49:02 & 0.8385 && 210 &\\
J1825$-$31 & 18:25:58 & $-$31:02:20 & 2.382 && 120 &\\
&\\
%J1835$-$09 & 18:35:22 & $-$09:27:34 & 0.2362 && 537 &\\ % 
% - found in lowlat. It was CONFIRMED in medlat
J1837$-$08 & 18:37:43 & $-$08:20:04 & 1.099 && 510 &\\
J1840$-$04 & 18:40:49 & $-$04:38:27& 0.4223 && 380 &\\
%J1855$-$16 & 18:55:21 & $-$16:21:25 & 1.703 && 19 &\\ % IS THIS ONE REAL????
J1900$-$09 & 19:00:14 & $-$09:28:07 & 1.424 && 150 &\\
J1902$-$10 & 19:02:18 & $-$10:39:33 & 0.7868 && 91 &\\
J1904$-$16 & 19:04:45 & $-$16:24:47 & 1.541 && 150 \\
\\
J1920$-$09 & 19:20:49 & $-$09:46:27 & 1.038 && 93 \\
%J2359$-$6537 & 23:59:15 & $-$65:40:32 & 0.6075 & 13 \\ - removed - hilat
\bottomrule
\end{tabular}
\label{table:partsolns}
\end{table}

\subsection{Discussion}\label{discussion1}
The mid-latitude portion of the HTRU survey has discovered 75 normal pulsars. There have also been several discoveries of MSPs \citep{batesmsps, keith2011}, and the discovery of a radio magnetar PSR~J1622--4950 \citep{levin2010}.

The addition of these pulsars alone will not contribute greatly to statistics about the population of pulsars. However, previous surveys of the Galactic plane extending to $|b| \leq 15\degree$ have had uneven coverage; multi-beam surveys by \citet[e.g.][]{mlc+01} and \citet{ebsb01} used integration times of 2100~s and 265~s respectively and did not cover the full area. We have now completed a survey of this region with uniform sensitivity, which will enable more precise study of the distribution of pulsars as a function of Galactic latitude. We have re-detected many previously-known pulsars in the survey region using the processing pipeline, which are briefly discussed in Appendix~\ref{sec:app}.

Despite the large number of discoveries the HTRU mid-latitude survey is yet to discover a young pulsar ($\tau_\mathrm{c} < 100~\mathrm{Kyr}$, $P<1~\mathrm{s}$). This, however, can be explained easily; the young pulsars are distributed along the Galactic plane at latitudes less than $3\degree$, a region of the sky which has already been observed to a limiting flux density of $0.15~\mathrm{mJy}$ in the PMPS \citep{mlc+01}. As the limiting flux density of the mid-latitude HTRU survey is $0.2~\mathrm{mJy}$, we would not expect to detect any such pulsars. The deep low-latitude part of the HTRU survey \citep[described in][]{keith2010}, however, should discover more young pulsars in the Galactic plane due to its improved sensitivity compared to the PMPS.

The Large Area Telescope on board the \textit{Fermi} Gamma-Ray Space Telescope has so far discovered many unassociated gamma-ray sources which were later found to be radio pulsars in targeted searches \citep[e.g.][]{ransom2011, keith2011b, cognard2011}. Gamma-ray pulsations from many previously-known pulsars were also detected by \textit{Fermi} \citep[e.g.][]{ray2010}. The standard metric for for the likelihood of pulsar being detected by Fermi is $\log(\sqrt{\dot{E}}/d^2)$ \citep[for a spin-down energy loss, $\dot{E}$, measured in erg~s$^{-1}$ and distance, $d$, in kpc; see\,][]{fermipsrcat}. For the majority of pulsars detected by \textit{Fermi}, this metric is greater than $\sim 17$. In the case of PSR~J0807--5421, $\log(\sqrt{\dot{E}}/d^2) = 17.2$, indicating that this pulsar is a candidate for detection in the \textit{Fermi} data. The other pulsars presented here fall below this threshold, and seem unlikely to be detected by \textit{Fermi}, however, there is a large uncertainty in the distance estimated from the Galactic electron distribution model \citep{ne2001}. If we assume that the distances are over-estimated by a factor of two, and recompute the metric, PSR~J0807--5421 remains the only source to satisfy $\log(\sqrt{\dot{E}}/d^2) > 17$. 

\section{Implementing an Artificial Neural Network}
\subsection{Overview of computer learning}
\begin{figure}
	\begin{center}
		\includegraphics[width=8cm]{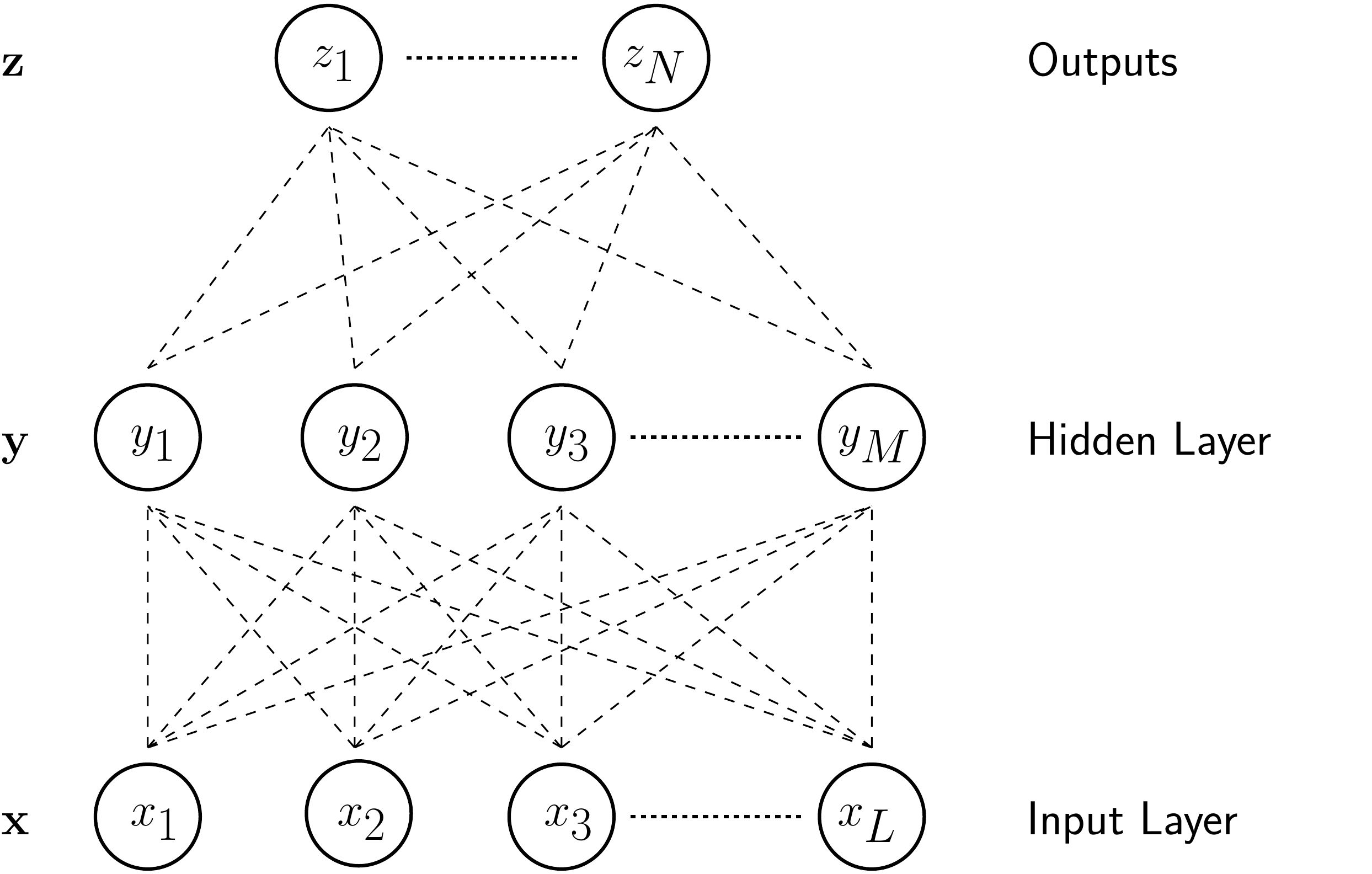}
	\end{center}
	\caption{Schematic diagram of an ANN, showing the input layer, $\mathbf{x}$, one of the `hidden layers', $\mathbf{y}$, and the output layer, $\mathbf{z}$.}
	\label{fig:NN}
\end{figure}

An Artificial Neural Network (ANN) is best described in terms of layers of `neurons', or units --- one-dimensional matrices --- where each unit is connected to every unit in the layers above and below it (see Figure~\ref{fig:NN}). In this scheme, the bottom layer is known as the `input layer', the top known as the `output layer' and any layers between the two are conventionally known as `hidden layers'; the layers which make up the ANN need not contain the same number of units. Hence the matrices $\mathbf{x}$ and $\mathbf{y}$ in Figure~\ref{fig:NN} are
\[
 \mathbf{x} =
 \begin{bmatrix}
  x_{1} \\
  x_{2} \\
  \vdots \\
  x_{L}
 \end{bmatrix}
 ,\qquad
 \mathbf{y} =
 \begin{bmatrix}
  y_{1} \\
  y_{2} \\
  \vdots \\
  y_{M}
 \end{bmatrix}.
\]

The connections between the input layer, $\mathbf{x}$, and the second layer, $\mathbf{y}$, of the ANN are then a two-dimensional matrix of weights,
\[
 \mathbf{w} =
 \begin{bmatrix}
  w_{1,1} & w_{1,2} & \cdots & w_{1,L} \\
  w_{2,1} & w_{2,2} & \cdots & w_{2,L} \\
  \vdots  & \vdots  & \ddots & \vdots  \\
  w_{M,1} & w_{M,2} & \cdots & w_{M,L}
 \end{bmatrix},
\]
ensuring that the weight of the connection between $x_1$ and $y_1$ need not be the same as that between $x_1$ and $y_2$.

\begin{figure}
	\begin{center}	\includegraphics[width=8cm]{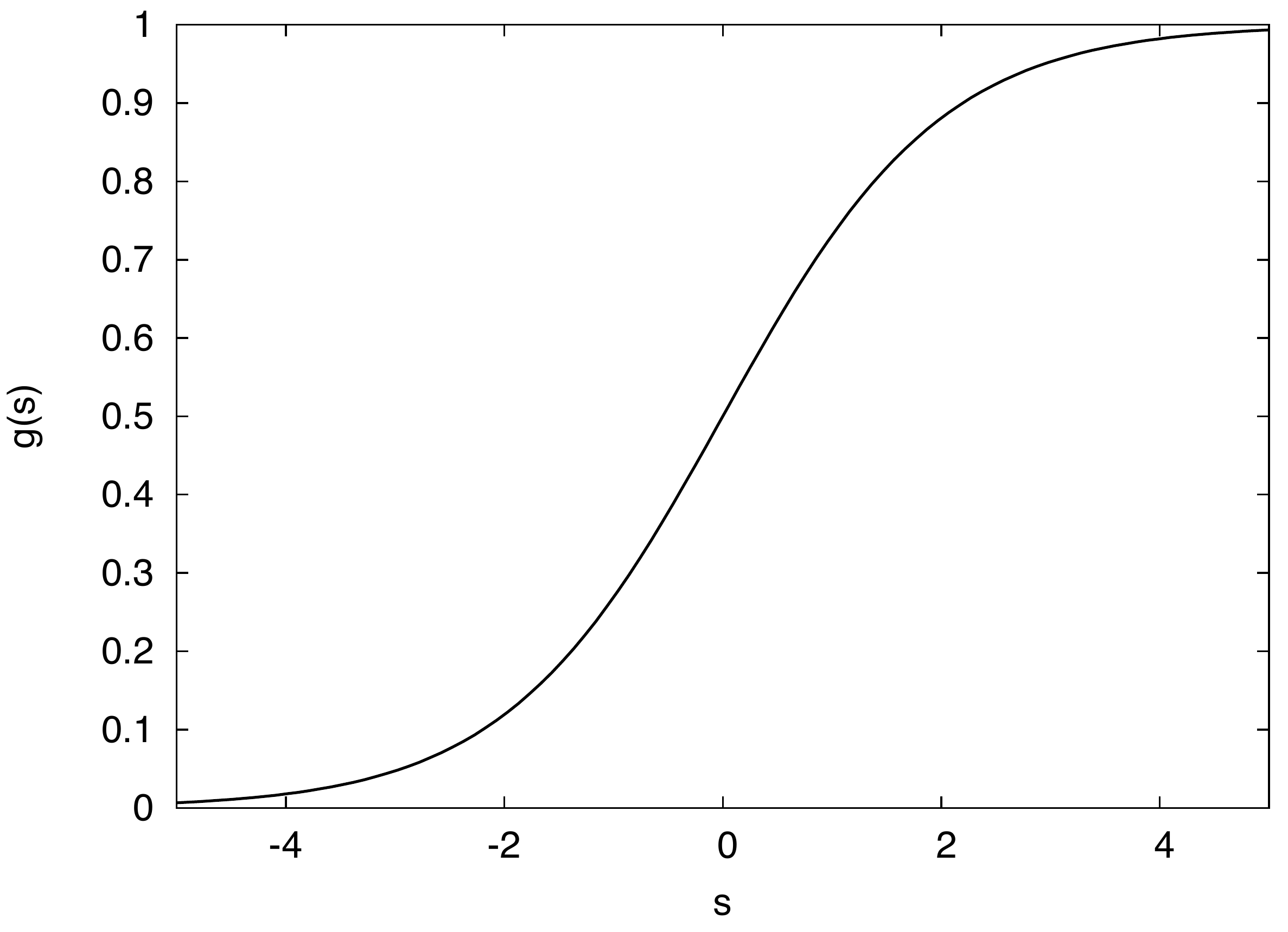}
	\end{center}
	\caption{Plot of the logistic sigmoid function, Equation~\ref{eq:NN_AF}. This function is useful because for small $s$, this can be used to approximate linear behaviour, but can also model non-linear behaviour in the general case.}
	\label{fig:AF}
\end{figure}

At each unit, $y_m$, the weighted sum of the layer below,
\begin{equation}
s_m=\sum_{l=1}^{L} w_{ml}x_l
\label{eq:NNsum}
\end{equation}
is calculated, before the calculation of $y_m$ using the activation function, 
\begin{equation}
y_m = g(s_m).
\end{equation}
The function $g(s_m)$ often takes the form
\begin{equation}
g(s_m) \equiv \frac{1}{1+\exp{(-s_m)}},
\label{eq:NN_AF}
\end{equation}
which is known as a `logistic sigmoid function' due to its shape (shown in Figure~\ref{fig:AF}), although any function may be used. For example, if $g(s_m) = s_m$, the ANN would only be able to reproduce linear functions, whereas by choosing a function of the form shown in Equation~(\ref{eq:NN_AF}), one allows for both non-linear (the general case) and linear behaviour (in the case of small $s$) of the input to be weighted \citep{looney}. Values then propagate through the network from the input layer up to the output layer. As with the input and hidden layers, the output layer can contain an arbitrary number of units; however, for most `simple' yes or no scenarios, two output values are sufficient (one signifying a `yes' score, the other a `no' score).

In order for the matrix $\textbf{w}$ to be populated, the ANN must be trained using a set of `patterns' (in this case, a set of scores which describe pulsar candidates) for which the desired output from the ANN is known. This collection of patterns is called a `training set'.

A common algorithm for training ANNs is `back-propagation', which is described in detail in \citet{bishop}. A general overview, however, is as follows; with the weights, $\mathbf{w}$, set to some initial value, a pattern is passed to the input layer, $\mathbf{x}$. These numbers propagate through the ANN as described above to produce the output vector, $\mathbf{z}$, known as `forward propagation'.

The error function for each pattern (designated by $k$), $E_k$, may then be computed using a sum of squares method for output $z_k$ and desired output $t_k$ (the `target') as
\begin{equation}
E_k=\frac{1}{2}\sum_k (z_k - t_k)^2,
\end{equation}
and a total error function defined as
\begin{equation}
E=\sum_k E_k.
\label{eq:errf}
\end{equation}
The derivative of $E$ with respect to each of the weights in the ANN can be calculated, and used to repopulate the $\mathbf{w}$ matrix with improved values. By repeating this process a number of times, the error between the input pattern and the target is minimised, resulting in a fully trained ANN.

\subsection{\label{nettests}Tests used to generate ANN input scores}
In order to generate the patterns used for training and using the ANN, a series of scores have been developed to try and describe each candidate as fully as possible. They were developed as an advancement of work by \citet{kel+09} and \citet{eatough2010} and hence some scores from that work are included here. The scores are listed in Table~\ref{table:NNscores}, and discussed below.

\subsubsection{Candidate Parameters}
The first scores generated are the pulse period in milliseconds, the DM in $\mathrm{cm}^{-3}\,\mathrm{pc}$, and the signal-to-noise ratio of the detection. These are read directly from the candidate metadata, and are generated during the processing.

Other scores include the pulse width and the $\chi^2$ value from fitting the pulse profile with a sine function (discussed below). These might ordinarily discriminate against many MSPs which often have wide pulse duty cycles compared to the normal pulsars \citep[e.g.][]{kramer1998}. By including pulse period as a score, it was hoped this would not be the case. Similarly, including the DM should prevent highly scattered pulsars being given a low ranking and terrestrial signals being ranked highly.

\begin{figure*}
	\begin{center}	\includegraphics[width=17.5cm]{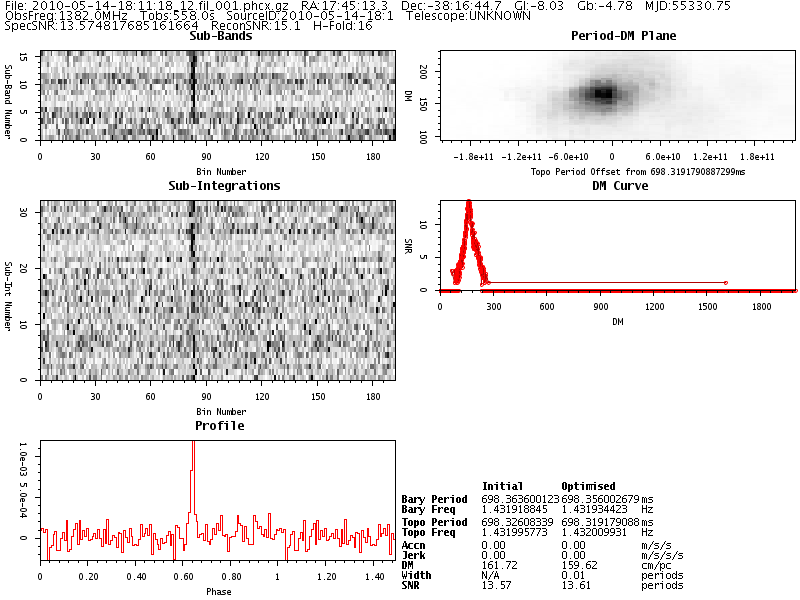}
	\end{center}
	\caption{Candidate plot for PSR J1745$-$3812, showing features typical of a good candidate pulsar. Starting from the top-right plot, and moving clockwise, the plots represent the S/N ratio of the source as a function of folding period and DM; S/N ratio as a function of DM; the folded pulse profile at the best values of period and DM; the folded pulse profile as a function of observing frequency; and the folded pulse profile as a function of observing time.
}
	\label{fig:cand}
\end{figure*}

\subsubsection{Profile Fitting}\label{sec:profilefitting}
To test for extremely wide pulse profiles, $\sin$ and $\sin^2$ functions are fitted to the pulse profile. Often such wide profiles are indicative of radio frequency interference (RFI), which can be mistakenly identified by the processing pipeline as a candidate. Therefore, we might expect a high $\chi^2$ value to indicate a pulsar.

To test for a ``typical'' profile shape, a single and double gaussian function are also fitted. Therefore, ignoring scattering which is often not important at 1.4~GHz, a low $\chi^2$ value is expected to indicate a pulsar. The FWHM of the gaussian, and alternative measurements of how well the gaussian fits the data, are also passed as scores.

Finally, the profile is tested to see how well it can be described as noise. A histogram is made of the values in the pulse profile, and is fitted with a Gaussian. The position of the peak of this Gaussian is passed as a score, as is the ratio of the amplitudes of the histogram to the fitted Gaussian. The histogram of RFI which has a noise-like profile is expected to be well described by a Gaussian centred on zero, whereas other profile shapes will cause the distribution of values to be skewed, and not described by a Gaussian.

A histogram of the first derivative of the pulse profile is also fitted with a Gaussian, and the offset from the pulse profile histogram is passed as a score. This fit will peak near zero in the case of a noise-like profile or a Gaussian-like profile, but for some signals (e.g.\,saw-tooth pulses), this will not be the case.

To complete the description of the pulse profile, we compute the number of distinct maxima in the pulse profile and pass that as a score. We then calculate the mean amplitude across all phase bins of the pulse profile, and subtract this from the original profile. The result is then integrated to compute the area, which is used as a score, which discriminates between different pulse widths and shapes.

\subsubsection{Dispersion Measure Response}\label{sec:dmresponse}
The signal-to-noise (S/N) ratio of the signal as a function of trial DM is recorded for each candidate during the data processing (the ``DM curve''). Dedispersion at an incorrect DM will cause a pulse to be smeared by an amount $\Delta\tau$ (in seconds), given by
\begin{equation}
\Delta\tau = 8.3 \times 10^3 \mathrm{DM} \nu_{\mathrm{MHz}}^{-3} \Delta\nu\label{dispersion}
\end{equation}
across an observing bandwidth of $\Delta\nu$ which is centred at frequency $\nu$, where both frequencies are in units of MHz. The S/N ratio of a pulse with effective width $W_{\rm{eff}}$ and period $P$ varies as 
\begin{equation}
\rm{S/N} \propto \sqrt{\frac{P - W_\mathrm{eff} }{ W_\mathrm{eff} } },
\label{eq:dewey1}
\end{equation}
and so the smearing of the pulse causes a variation of the S/N ratio (see the middle right-hand panel in Figure~\ref{fig:cand}). We fit this relationship to the data and record the $\chi^2$ of the fit, and the shift in best DM as scores for the ANN.

If we rearrange Equation~\ref{eq:dewey1} in terms of the flux density,
\begin{equation}
S_{\rm min} = k \sqrt{\frac{W_{\mathrm{eff}}}{P - W_{\mathrm{eff}}}}\label{senseq}
\end{equation}
we can group all system-dependant parameters into a single constant of proportionality, $k$. To create another score for the ANN, we calculate the value of $k$ for the DM curve data, and for the best fit. In the ideal case of a pulsar, these two values would be equal, and they are both used as scores in the ANN.

\subsubsection{Frequency Sub-band Data}
The candidate plot (Figure~\ref{fig:cand}) shows the folded pulse profile as a function of observing frequency, in a set of frequency sub-bands across the observing bandwidth. As broadband radio-emitting objects, a pulsar is expected to be visible right across the observing bandwidth, whereas RFI can often occur as a narrow-band phenomenon, and only be visible in one or two of the frequency sub-bands.

To test this, we perform three tests on this plot.
\begin{itemize}
\item First, the standard deviation of the peak bin in each sub-band is calculated, normalised to the width of the pulse. For a broadband signal, the standard deviation should be small.
\item We then calculate the mean of the correlation coefficient of each sub-band with the folded pulse profile. For narrow-band signals, indicative of RFI, only one or two of the sub-bands will correlate strongly with the pulse profile.
\item Finally, the correlation coefficient is calculated for all pairs of sub-bands. For a strong broadband signal, again the mean correlation coefficient will be high.
\end{itemize}

While a set of similar tests could be implemented for the sub-integration data (pulse profile as a function of time through the observation), it was decided not to include them in this ANN. Such tests should select against pulsars in short-period binary systems (where the pulsar's motion causes the pulses not to fall in a straight line in this plot), and also against nulling pulsars (and, potentially, bright RRATs) where the pulse profile might appear and disappear as a function of time. 

\begin{table}
	\begin{center}
	\caption{List of individual scores used as input to the ANN, and the average correlation between that score and the ANN ``Y" output (see text).}
		\begin{tabular}{llr}
		\toprule
		\# & Description of score & $\rho_\mathrm{SY}$\\
		\midrule
		\multicolumn{3}{l}{Candidate Parameters}\\
		%\midrule
		1 & Best period (ms) & $-0.09$\\
		2 & Best DM value, DM$_\mathrm{best}$ & $-0.15$\\
		3 & Best S/N ratio & 0.02\\%$^\star$  \\
		4 & Pulse width & $-0.30$\\%$^\star$\\
		\midrule
		\multicolumn{3}{l}{Sinusoid Fitting}\\
		%\midrule
		5 & $\chi^2$ value: fitting pulse profile with a $\sin$ curve & $0.52$\\
		6 & $\chi^2$ value: fitting pulse profile with a $\sin^2$ curve & 0.02\\
		\midrule
		\multicolumn{3}{l}{Gaussian Fitting}\\
		%\midrule
		7 & $\chi^2$ value: fitting profile with Gaussian & $-0.45$\\
		8 & FHWM of Gaussian fit & $-0.08$\\
		9 & $\chi^2$ value: fitting profile with two Gaussians & $-0.62$ \\
		10 & Mean FHWM from fitting profile &\\
		& \hspace{2mm}with two Gaussians & $-0.11$\\
		\midrule
		\multicolumn{3}{l}{Profile Histogram Tests}\\
		11 & Offset of profile histogram from zero & $0.28$\\
		12 & Max.\,of profile histogram / &\\
		&\hspace{2mm} Max.\,of fitted Gaussian & $-0.04$\\ 
		13& Histogram of d(profile)/d$x$, &\\
		&\hspace{2mm}find offset from score 11 & $-0.32$\\
		\midrule
		\multicolumn{3}{l}{DM Curve Fitting}\\
		%\midrule
		14 & S/N$_\mathrm{data}$ / $\sqrt{(P-W)/W}$ & 0.01\\
		15 & S/N$_\mathrm{fit}$ / $\sqrt{(P-W)/W}$ & $-0.28$\\
		16 & mod(DM$_\mathrm{fit}$ - DM$_\mathrm{best}$) & $-0.23$\\
		17 & $\chi^2$ value: DM curve fit & $-0.47$\\%$^\star$\\
		\midrule
		\multicolumn{3}{l}{Sub-band Tests}\\
		%\midrule
		18 & RMS of peak positions in all sub-bands & 0.03\\%$^\star$\\
		19 & Average correlation coeff.\,&\\
		& \hspace{2mm}for each pair of sub-bands & 0.28\\%$^\star$\\
		20 & Sum of correlation coefficients & 0.35\\
		\midrule
		\multicolumn{3}{l}{Pulse Profile Tests}\\
		%\midrule		
		21 & Number of peaks in the pulse profile & $-0.51$\\
		22 & Area under the pulse profile & \\
		& \hspace{2mm}after subtracting mean & $0.55$\\
		\bottomrule
		\end{tabular}
		\label{table:NNscores}
	\end{center}
%	$^\star$Indicates the test originated in the work of \citeauthor{eatough2010}
\end{table}
% Did testing for this table here:
%/Users/sbates/Documents/Physics/Pulsars/hitrun/NN/2009-06-30/final/22scores/100_1_0.9982_0.0160

\subsection{Applying the ANN to data from the HTRU survey}
\subsubsection{Training}\label{sec:training}
Having decided upon a set of scores to describe the candidates, an ANN was trained and generated using the Stuttgart Neural Network Simulator\footnote{http://www.ra.cs.uni-tuebingen.de/SNNS/}. 
The scores detailed in Section \ref{nettests} were generated for a selection of initial HTRU data which contained 70 pulsar and 200 non-pulsar candidate files, picked at random from the data (this training set was so small because the ANN was first implemented early-on in the data-taking process, when few known pulsars had been observed). These were divided between a `training set' and a `validation set', and each file was given a `target', that is, the desired output from the ANN, either ``1 0" for pulsars, and ``0 1" for non-pulsars. 

Following \citet{eatough2010}, the ANN was set up as a 22:22:2 (22 units in the input and hidden layers, and 2 in the output layer), and weights were initially randomised. Training was performed using the training set, with the validation set used as an independent check of the error (Equation \ref{eq:errf}).

As training progresses, the error in the validation set gradually decreases, but eventually reaches a minimum, after which the error begins to rise. This is due to the ANN becoming `over-trained', and sensitive to specific properties of the training set. Therefore, optimum training is achieved when the validation error reaches the minimum point.

\subsubsection{Practical use}
A modification to the HTRU processing pipeline (\textsc{hitrun}, described in \citet{keith2010}), was made to pass candidates into the ANN. Although all candidates were kept for a more detailed inspection using an interactive interface, the ANN output was used to make a subset of the candidates for a quick inspection. Given the output format of ``X Y" (see Section \ref{sec:training}), candidates were rejected where $\rm{X} < 0.5$ and $\rm{Y} > 0.5$. This removed $\sim 99.7~\%$ of candidates, leaving a manageable number to be inspected by eye as data were processed.

For example, a typical data LTO-4 data tape would contain $\sim 350$ observations, each producing 150 candidates after processing. By using the ANN, the number of candidates to view is reduced to $\sim150$. After the previously-known pulsars are removed from this list (to avoid time being wasted on misidentification), this small number of candidates can be viewed very quickly.

\subsection{Analysis of the ANN}
After using the ANN for over a year, and with two years of data from the HTRU survey, we have obtained candidate files for 580 known pulsars (including those used in the training process), and are able to make a thorough analysis of the performance of the ANN with these data. The ANN was used to classify both MSPs and normal pulsars.

\subsubsection{Overall performance}
First, we look at the simplest, and in many ways the most important, metric of how well the ANN performs; what fraction of pulsars are detected. Before performing this analysis, all the pulsars in the training set were analysed separately to see how much bias they would cause on our results if they were included. Of the 70 pulsars in the training set, all 70 were identified as pulsars by the ANN. Therefore, while the ANN had clearly converged on weights suitable for the training set, these candidates were excluded from the rest of the analysis.

After removing the training pulsars, this left a set of 510 candidate files which each contained observations of a known pulsar. The ANN was able to correctly identify 85\% of these candidates as pulsars, which is a promising fraction. However, compared to 92\% in the work of \citeauthor{eatough2010}, this number seems a little disappointing. It is possible that this difference can be explained by two factors,
\begin{inparaenum}[\itshape a\upshape)]
\item the test set used in the analysis of \citeauthor{eatough2010} included pulsars used in the training set (Eatough, priv.\,comm.); and
\item \citeauthor{eatough2010} showed the strong dependance of an ANN's efficiency on pulsar parameters. The fraction detected will, therefore, strongly depend on the pulsars which make up the test set.
\end{inparaenum}

In the following sections, results from the ANN are studied in more detail. This will allow us to draw conclusions about the ability of an ANN to identify pulsars, and the necessary future work to improve these tools.

\subsubsection{Distribution and correlation of scores and output}
\begin{figure}
	\begin{center}	\includegraphics[width=8cm]{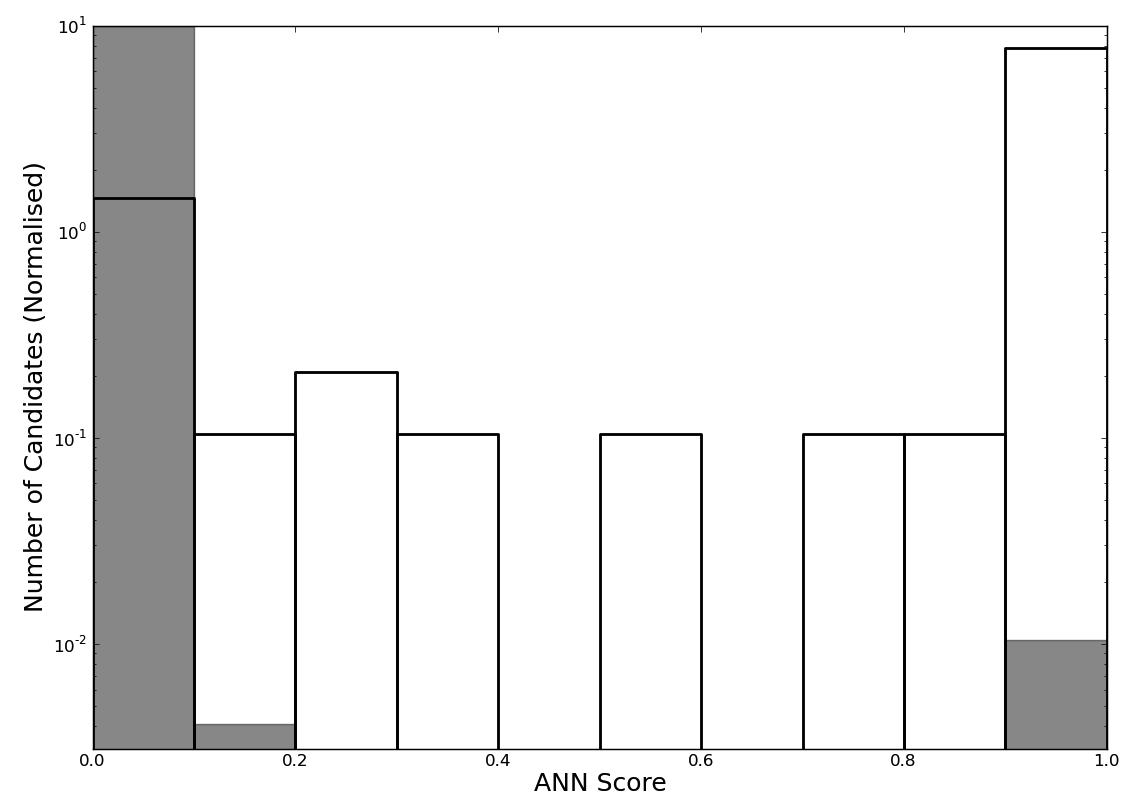}
	\end{center}
	\caption{Histogram of output `yes' scores from the ANN (in grey for all candidates, solid lines for known and newly-discovered pulsars only). From the overall sample of candidates, the vast majority are rejected by the ANN, but the majority of real pulsars are well-ranked by the ANN.}
	\label{fig:ANNoutput}
\end{figure}

Figure~\ref{fig:ANNoutput} shows (in grey) the distribution of output scores from the ANN for hundreds of thousands of candidate files chosen from \textsc{hitrun}, on a logarithmic y-axis. With only a small number of high `yes' scores, $\sim 99.7\%$ of candidates are rejected by the ANN. The solid lines in Figure~\ref{fig:ANNoutput} show the same scores but only for known or newly-discovered pulsars. Here, it can be seen that the majority of pulsars are detected by the ANN, as mentioned previously.

To test that the ANN was not creating contradictory output scores, the correlation coefficient, $\rho$, of the `yes' score, Y, with the `no' score, N, was calculated. One would naively expect $\rho_\mathrm{YN} \approx-1$ since the training set was composed entirely of candidates classed either as `pulsar' or `non-pulsar', and the targets used for training reflected this. The correlation coefficient was calculated to be $\rho_\mathrm{YN} = -0.9991$, confirming this hypothesis.

Correlation coefficient matrices were calculated for each of the scores in the input layer (shown in Table~\ref{table:NNscores}) with the `yes' and `no' scores. For each score parameter S, $\rho_\mathrm{S} = \rho_\mathrm{SY} \approx -\rho_\mathrm{SN}$, and hence all the inputs to the ANN cause the output scores to scale oppositely. The absolute value of $\rho_\mathrm{S}$ varies from 0.01 to 0.62 for different input scores, indicating that some scores are far more significant than others when the ANN produces the output.

From Table~\ref{table:NNscores}, we can see that there is a subsection of the scores which appear to dominate the output ratings. These are mainly the tests which evaluate the shape of the pulse profile (scores 5, 7, 9, 21 and 22), but also the $\chi^2$ from making a fit to the DM response curve (score 17), and the correlation coefficients for each sub-band with the pulse profile (score 20). These scores also scale in an intuitive way --- for example, when the DM curve fits well (lower $\chi^2$), then the ANN score is higher; and when the pulse profile is well correlated with the sub-band information, the ANN score tends to be increased.

\subsubsection{Output score as a function of pulse period}
\citeauthor{eatough2010} noted that the ANN used in their analysis was only able to detect $\sim 50\%$ of the pulsars with spin periods below 10~ms (not accounting for training set pulsars included in their sample). Our ANN has slightly improved this figure, recovering 65\% of pulsars in this category. 

\begin{figure*}
	\begin{center}	
		\includegraphics[width=5.8cm]{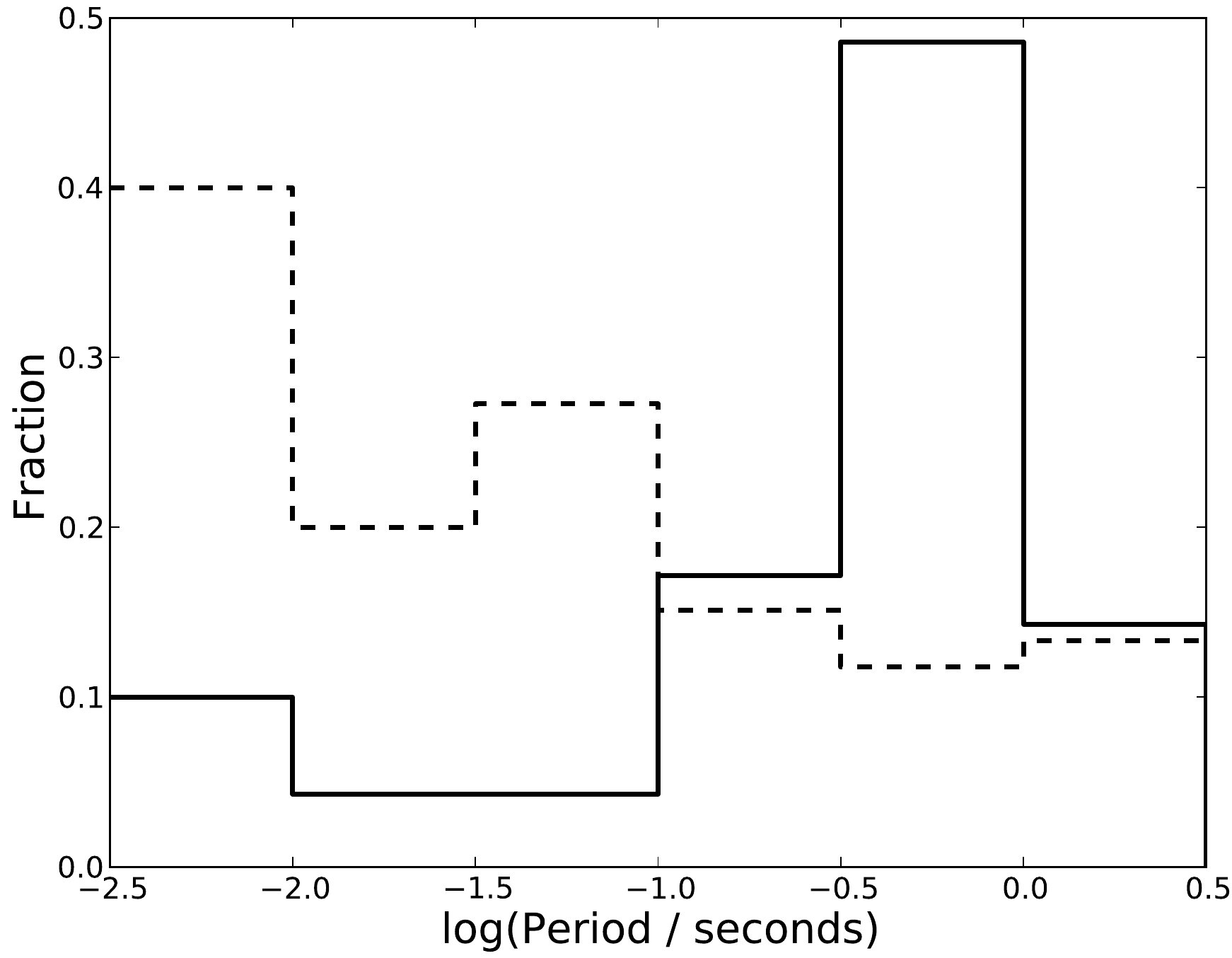}
		\includegraphics[width=5.8cm]{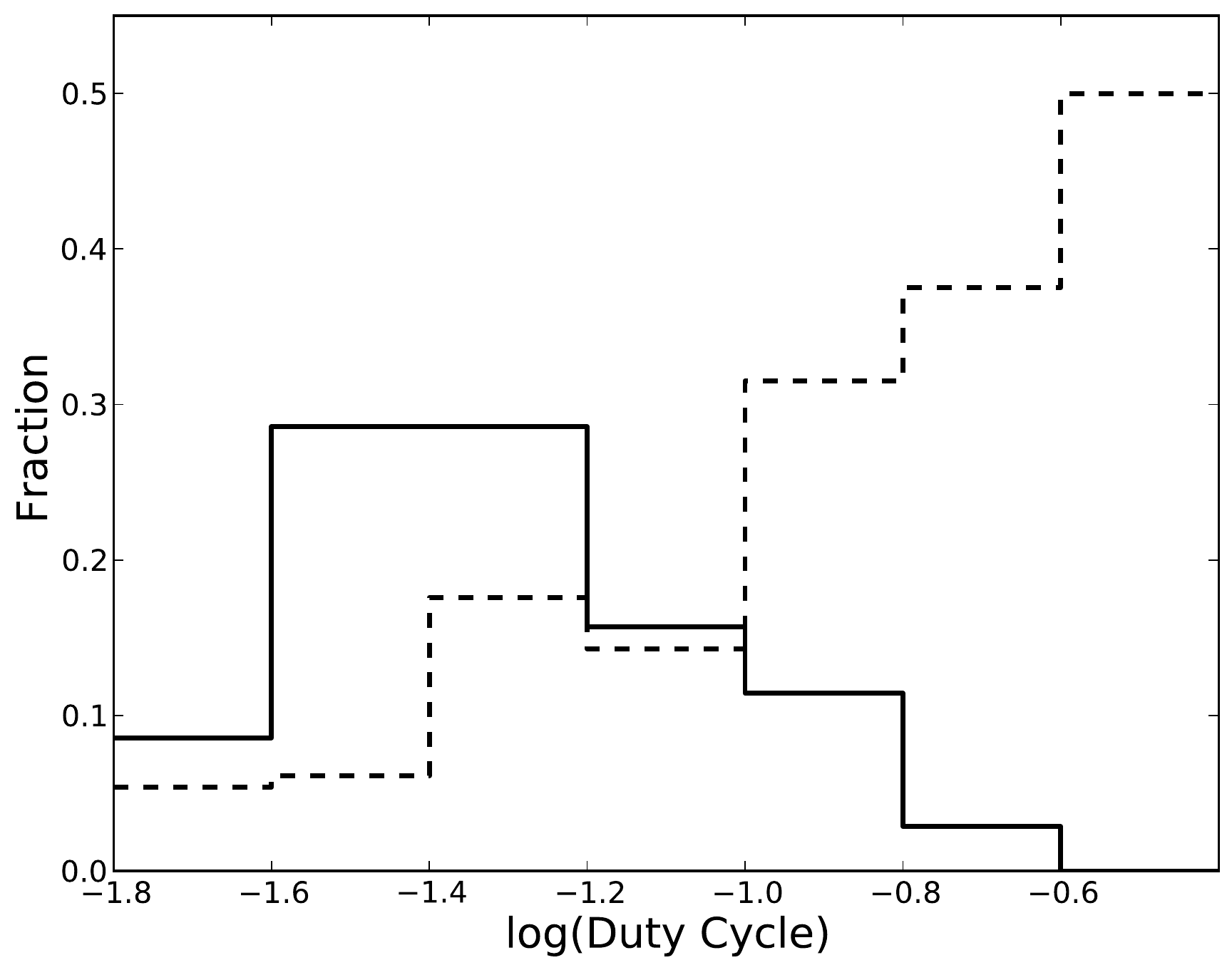}
		\includegraphics[width=5.8cm]{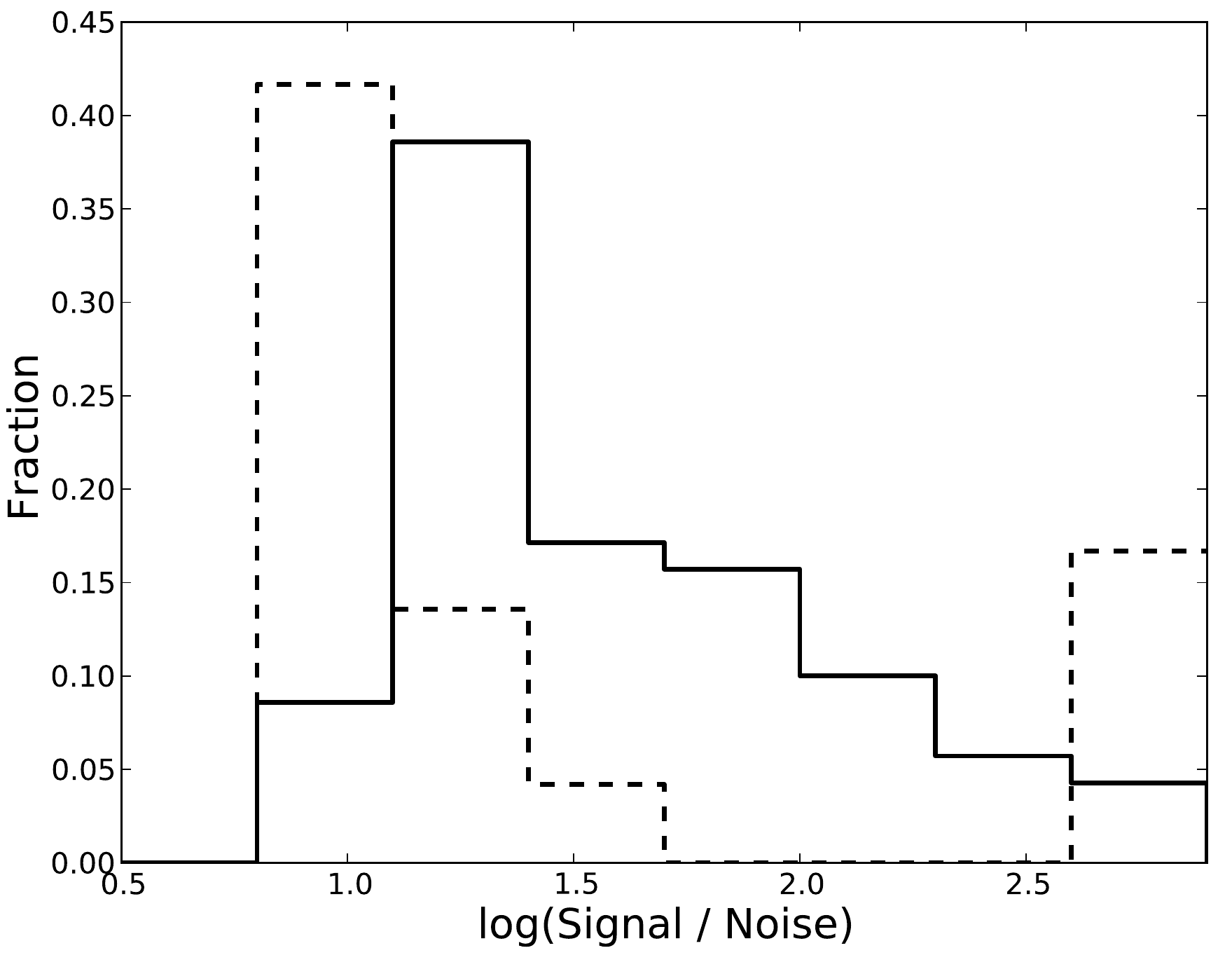}
	\end{center}
	\caption{The fraction of pulsars  that were undetected by the ANN (dashed lines) as a function of pulse period (left), pulse duty cycle (centre) and S/N ratio (right). The solid lines show what fraction of the training set was made up by pulsars with the corresponding property.}
	\label{fig:NNhist}
\end{figure*}

The fraction of pulsars detected at all pulse periods can be seen in Figure~\ref{fig:NNhist}. At pulse periods greater than 100~ms, the ANN performs well, detecting 86.2\% of the pulsars; at periods below 100~ms, the detection rate is 71\%. Clearly, there is an improvement in the performance at longer periods. However, as the pulse period is only 1 of 22 input scores, it is unlikely to be the deciding factor. Rather, other properties of this population (e.g.\,the larger pulse duty cycle at shorter periods) are also important in the scoring.

\subsubsection{What other properties are causing pulsars to be missed?}\label{sec:otherprops}
Histograms of our sample of pulsars as a function of pulse duty cycle and S/N ratio in the observation, as well as the fraction that are not detected by the ANN, are plotted in Figure~\ref{fig:NNhist}. Also included in these figures are histograms showing the distribution of these properties in the training set, marked with a solid line.

In Figure~\ref{fig:NNhist}, it can been seen that the ANN performs badly for wide pulses, where the duty cycle is $\gtrsim 20\%$. For pulses narrower than this, it performs rather well. The training set, however, contains no pulsars whose pulse duty cycle is greater than $\sim 16\%$. The right-most panel shows a similar trend; the ANN performs poorly where the S/N ratio is low (as would be expected with human inspection), but we can see that the training set contained few pulsars with a S/N ratio less than 15.

Our ANN is shown to be less effective at identifying short period and wide pulsars. Since the average duty cycle of MSPs is larger than that for the normal pulsars, in many cases this is simply a reflection of the difficulty of detecting MSPs, which have very narrow DM curves (see Section~\ref{sec:dmresponse}), are in a region of period space where there are many false candidates, and the detection is further complicated, in many cases, by binary motion.

However, the training process is the method by which a reliable set of weights in an ANN is created, and the ability of the ANN to identify pulsars is, therefore, dependant upon the training set that is used. While there are many intrinsic properties of MSPs which make their detection difficult, it might be that a training set comprised entirely of MSPs would produce better results. Further work on this, including the possibility of using simulated candidates for training purposes, are required before any strong conclusion can be drawn. 

That said, in the period when our ANN was first implemented at JBO, the majority of normal pulsars were discovered using this technique, and while the ANN is shown to be weaker at discovering MSPs, 3 were discovered this way.

Future improvements to such systems may include the need for separate ANNs for different classes of candidate. For example, an ANN trained specifically for narrow pulses, another for wide pulses, and potentially others for classifying fast binary systems or even RFI. 

\section{Conclusions}
In this paper we have presented 75 pulsars discovered in the mid-latitude portion of the HTRU survey. Further discoveries in that survey, including the low-latitude and all-sky portions, are sure to continue as more advanced processing techniques are applied to the data. While the main objective of the survey is the discovery of rapidly-rotating MSPs, many of the new discoveries will also be normal pulsars. As in the case of PSR~J1054--5946, some of these pulsars will display unusual behaviour and will enable further studies of the pulsar population including their origins and birth, their evolution, and their emission mechanism.

Current techniques in pulsar surveys tend to produce enormous numbers of candidates which must be sifted through to find targets for confirmation observations. While the application of ANNs has not proven to be a panacea for this problem, we have demonstrated that even a rudimentary ANN can provide an excellent way to quickly identify an initial group of candidates before a more time-consuming approach is required, using the traditional techniques. It is also only by this approach that every single candidate will be, in some sense, ``looked at'', regardless of signal-to-noise ratio or other artificial cut-offs. As future pulsars surveys by instruments such as LOFAR \citep{bws} and the SKA \citep{smits2008} produce even larger volumes of candidates, such techniques will become increasingly important.

In this paper we have seen that our ANN is capable of detecting pulsars at all pulse periods, but is appreciably less adept at identifying strong candidates with a large pulse duty cycle, and with millisecond periods. Given that we estimate the ANN detected pulsars with an accuracy of $\sim85\%$, we would estimate that for the mid-latitude dataset, $\sim 15$ normal pulsars might be present in the data that were not detected by the ANN, and were missed by other means. However, the ANN was used as a complementary technique; short period candidates, and many with longer periods, were also looked at by eye, due to the known shortcomings. The ANN was also only implemented at one of our processing sites, and so we expect that this estimate serves as an upper limit.

Further work on this technique is required, in order to see how much improvement can be made on the detection of MSPs and the training process itself, but nevertheless this technique is shown to work. It should be remembered, however, that in order to maximise the possibility of serendipitous discoveries, human inspection of candidates, at some level, is still required.

\section*{ACKNOWLEDGEMENTS}
The authors thank Cristobal Espinoza for his helpful comments and expertise on pulsar glitches. The Parkes Observatory is part of the Australia Telescope which is funded by the Commonwealth of Australia for operation as a National Facility managed by CSIRO. We are grateful to the anonymous referee for helpful comments which have improved the quality of this work.
\newpage

\bibliography{allrefs}

\begin{thebibliography}{}

\bibitem[\protect\citeauthoryear{{Abdo} et~al.}{{Abdo}
  et~al.}{2010}]{fermipsrcat}
{Abdo} A.~A. et~al., 2010, ApJS, 187, 460

\bibitem[\protect\citeauthoryear{{Bailes} et~al.}{{Bailes}
  et~al.}{2011}]{bailesplanet}
{Bailes} M. et~al., 2011, Science, 333, 1717

\bibitem[\protect\citeauthoryear{{Bates} et~al.}{{Bates}
  et~al.}{2011}]{batesmsps}
{Bates} S.~D. et~al., 2011, MNRAS, 416, 2455

\bibitem[\protect\citeauthoryear{{Bhat} et~al.}{{Bhat} et~al.}{2004}]{bcc+04}
{Bhat} N.~D.~R., {Cordes} J.~M., {Camilo} F., {Nice} D.~J.,  {Lorimer} D.~R.,
  2004, ApJ, 605, 759

\bibitem[\protect\citeauthoryear{{Bishop}}{{Bishop}}{1995}]{bishop}
{Bishop} C.~M., 1995, {Neural Networks for Pattern Recognition}.
\newblock Oxford University Press

\bibitem[\protect\citeauthoryear{{Boyles} et~al.}{{Boyles}
  et~al.}{2010}]{boyles2010}
{Boyles} J. et~al., 2010, in Bulletin of the American Astronomical Society,
  Vol.~42, Bulletin of the American Astronomical Society, p. 464

\bibitem[\protect\citeauthoryear{{Burke-Spolaor} et~al.}{{Burke-Spolaor}
  et~al.}{2011}]{htru3}
{Burke-Spolaor} S. et~al., 2011, MNRAS, 416, 2465

\bibitem[\protect\citeauthoryear{{Camilo} et~al.}{{Camilo}
  et~al.}{2012}]{camilo2012}
{Camilo} F., {Ransom} S.~M., {Chatterjee} S., {Johnston} S.,  {Demorest} P.,
  2012, ApJ, 746, 63

\bibitem[\protect\citeauthoryear{{Cognard} et~al.}{{Cognard}
  et~al.}{2011}]{cognard2011}
{Cognard} I. et~al., 2011, ApJ, 732, 47

\bibitem[\protect\citeauthoryear{{Cordes} \& {Lazio}}{{Cordes} \&
  {Lazio}}{2002}]{ne2001}
{Cordes} J.~M.,  {Lazio} T.~J.~W., 2002, ArXiv Astrophysics e-prints
  (astro-ph/0207156)

\bibitem[\protect\citeauthoryear{Eatough}{Eatough}{2009}]{eatoughthesis}
Eatough R.~P., 2009, Ph.D. thesis, {University of Manchester}

\bibitem[\protect\citeauthoryear{{Eatough} et~al.}{{Eatough}
  et~al.}{2010}]{eatough2010}
{Eatough} R.~P., {Molkenthin} N., {Kramer} M., {Noutsos} A., {Keith} M.~J.,
  {Stappers} B.~W.,  {Lyne} A.~G., 2010, MNRAS, 407, 2443

\bibitem[\protect\citeauthoryear{{Edwards} et~al.}{{Edwards}
  et~al.}{2001}]{ebsb01}
{Edwards} R.~T., {Bailes} M., {van Straten} W.,  {Britton} M.~C., 2001, MNRAS,
  326, 358

\bibitem[\protect\citeauthoryear{{Espinoza} et~al.}{{Espinoza}
  et~al.}{2011}]{espinoza2011}
{Espinoza} C.~M., {Lyne} A.~G., {Stappers} B.~W.,  {Kramer} M., 2011, MNRAS,
  414, 1679

\bibitem[\protect\citeauthoryear{Faucher-Gigu{\`e}re \&
  Kaspi}{Faucher-Gigu{\`e}re \& Kaspi}{2006}]{fk06}
Faucher-Gigu{\`e}re C.~A.,  Kaspi V.~M., 2006, ApJ, 643, 332, in press

\bibitem[\protect\citeauthoryear{{Faucher-Gigu{\`e}re} \&
  {Loeb}}{{Faucher-Gigu{\`e}re} \& {Loeb}}{2011}]{fgl2010}
{Faucher-Gigu{\`e}re} C.-A.,  {Loeb} A., 2011, MNRAS, 415, 3951

\bibitem[\protect\citeauthoryear{{Ferdman} et~al.}{{Ferdman}
  et~al.}{2010}]{ferdman2010}
{Ferdman} R.~D. et~al., 2010, Classical and Quantum Gravity, 27, 084014

\bibitem[\protect\citeauthoryear{{Hobbs} et~al.}{{Hobbs}
  et~al.}{2009}]{hobbs2009}
{Hobbs} G.~B. et~al., 2009, Publ. Astr. Soc. Aust., 26, 103

\bibitem[\protect\citeauthoryear{{Hobbs}, {Edwards}, \& {Manchester}}{{Hobbs}
  et~al.}{2006}]{tempo2_1}
{Hobbs} G.~B., {Edwards} R.~T.,  {Manchester} R.~N., 2006, MNRAS, 369, 655

\bibitem[\protect\citeauthoryear{{Jenet} et~al.}{{Jenet}
  et~al.}{2009}]{jenet2009}
{Jenet} F. et~al., 2009, ArXiv Astrophysics e-prints (astro-ph/0909.1058)

\bibitem[\protect\citeauthoryear{{Keane} \& {Kramer}}{{Keane} \&
  {Kramer}}{2008}]{keane2008}
{Keane} E.~F.,  {Kramer} M., 2008, MNRAS, 391, 2009

\bibitem[\protect\citeauthoryear{{Keith} et~al.}{{Keith} et~al.}{2009}]{kel+09}
{Keith} M.~J., {Eatough} R.~P., {Lyne} A.~G., {Kramer} M., {Possenti} A.,
  {Camilo} F.,  {Manchester} R.~N., 2009, MNRAS, 395, 837

\bibitem[\protect\citeauthoryear{{Keith} et~al.}{{Keith}
  et~al.}{2010}]{keith2010}
{Keith} M.~J. et~al., 2010, MNRAS, 409, 619

\bibitem[\protect\citeauthoryear{{Keith} et~al.}{{Keith}
  et~al.}{2012}]{keith2011}
{Keith} M.~J. et~al., 2012, MNRAS, 419, 1752

\bibitem[\protect\citeauthoryear{{Keith} et~al.}{{Keith}
  et~al.}{2011}]{keith2011b}
{Keith} M.~J. et~al., 2011, MNRAS, 414, 1292

\bibitem[\protect\citeauthoryear{{Kramer} et~al.}{{Kramer}
  et~al.}{2004}]{kramer2004}
{Kramer} M., {Backer} D.~C., {Cordes} J.~M., {Lazio} T.~J.~W., {Stappers}
  B.~W.,  {Johnston} S., 2004, New Astronomy Review, 48, 993

\bibitem[\protect\citeauthoryear{{Kramer} et~al.}{{Kramer}
  et~al.}{2006}]{kramer2006a}
{Kramer} M., {Lyne} A.~G., {O'Brien} J.~T., {Jordan} C.~A.,  {Lorimer} D.~R.,
  2006, Science, 312, 549

\bibitem[\protect\citeauthoryear{{Kramer} et~al.}{{Kramer}
  et~al.}{1998}]{kramer1998}
{Kramer} M., {Xilouris} K.~M., {Lorimer} D.~R., {Doroshenko} O., {Jessner} A.,
  {Wielebinski} R., {Wolszczan} A.,  {Camilo} F., 1998, ApJ, 501, 270

\bibitem[\protect\citeauthoryear{{Levin} et~al.}{{Levin}
  et~al.}{2010}]{levin2010}
{Levin} L. et~al., 2010, ApJL, 721, L33

\bibitem[\protect\citeauthoryear{{Looney}}{{Looney}}{1997}]{looney}
{Looney} C.~G., 1997, {Pattern Recognition Using Neural Networks}.
\newblock Oxford University Press

\bibitem[\protect\citeauthoryear{{Lorimer}}{{Lorimer}}{2010}]{lorimer2010}
{Lorimer} D.~R., 2010, ArXiv Astrophysics e-prints (astro-ph/1008.1928)

\bibitem[\protect\citeauthoryear{{Lyne} et~al.}{{Lyne} et~al.}{2010}]{lyne2010}
{Lyne} A., {Hobbs} G., {Kramer} M., {Stairs} I.,  {Stappers} B., 2010, Science,
  329, 408

\bibitem[\protect\citeauthoryear{Lyne, Manchester, \& Taylor}{Lyne
  et~al.}{1985}]{lmt85}
Lyne A.~G., Manchester R.~N.,  Taylor J.~H., 1985, MNRAS, 213, 613

\bibitem[\protect\citeauthoryear{{Lyne} \& {Smith}}{{Lyne} \&
  {Smith}}{2005}]{lynesmith}
{Lyne} A.~G.,  {Smith} F.~G., 2005, {Pulsar Astronomy}

\bibitem[\protect\citeauthoryear{Manchester et~al.}{Manchester
  et~al.}{2005}]{mhth05}
Manchester R.~N., Hobbs G.~B., Teoh A.,  Hobbs M., 2005, AJ, 129, 1993

\bibitem[\protect\citeauthoryear{Manchester et~al.}{Manchester
  et~al.}{2001}]{mlc+01}
Manchester R.~N. et~al., 2001, MNRAS, 328, 17

\bibitem[\protect\citeauthoryear{{McLaughlin} et~al.}{{McLaughlin}
  et~al.}{2006}]{rrats2006}
{McLaughlin} M.~A. et~al., 2006, Nature, 439, 817

\bibitem[\protect\citeauthoryear{O'Brien}{O'Brien}{2008}]{obrienthesis}
O'Brien J.~T., 2008, Ph.D. thesis, {University of Manchester}

\bibitem[\protect\citeauthoryear{{Ransom} et~al.}{{Ransom}
  et~al.}{2011}]{ransom2011}
{Ransom} S.~M. et~al., 2011, ApJL, 727, L16

\bibitem[\protect\citeauthoryear{{Ray} \& {Saz Parkinson}}{{Ray} \& {Saz
  Parkinson}}{2010}]{ray2010}
{Ray} P.~S.,  {Saz Parkinson} P.~M., 2010, ArXiv Astrophysics e-prints
  (astro-ph/1007.2183)

\bibitem[\protect\citeauthoryear{Ridley \& {Lorimer}}{Ridley \&
  {Lorimer}}{2010}]{ridley2010}
Ridley J.~P.,  {Lorimer} D.~R., 2010, MNRAS, 404, 1081

\bibitem[\protect\citeauthoryear{{Rosen} et~al.}{{Rosen}
  et~al.}{2010}]{rosen2010}
{Rosen} R. et~al., 2010, Astronomy Education Review, 9, 010106

\bibitem[\protect\citeauthoryear{{Smits} et~al.}{{Smits}
  et~al.}{2009}]{smits2008}
{Smits} R., {Kramer} M., {Stappers} B., {Lorimer} D.~R., {Cordes} J.,
  {Faulkner} A., 2009, A\&A, 493, 1161

\bibitem[\protect\citeauthoryear{{Stappers} et~al.}{{Stappers}
  et~al.}{2011}]{stappers2011}
{Stappers} B.~W. et~al., 2011, A\&A, 530, A80

\bibitem[\protect\citeauthoryear{{Taylor} \& {Cordes}}{{Taylor} \&
  {Cordes}}{1993}]{taylor1993}
{Taylor} J.~H.,  {Cordes} J.~M., 1993, ApJ, 411, 674

\bibitem[\protect\citeauthoryear{{van Leeuwen} \& {Stappers}}{{van Leeuwen} \&
  {Stappers}}{2010}]{bws}
{van Leeuwen} J.,  {Stappers} B.~W., 2010, A\&A, 509, A7

\bibitem[\protect\citeauthoryear{{Weltevrede}, {Johnston}, \&
  {Espinoza}}{{Weltevrede} et~al.}{2011}]{weltevrede2011}
{Weltevrede} P., {Johnston} S.,  {Espinoza} C.~M., 2011, MNRAS, 411, 1917

\bibitem[\protect\citeauthoryear{{Weltevrede} et~al.}{{Weltevrede}
  et~al.}{2010}]{weltevrede2010}
{Weltevrede} P. et~al., 2010, Publications of the Astronomical Society of
  Australia, 27, 64

\end{thebibliography}
\bibliographystyle{mnras}

\appendix
\section[]{Details of the previously-known pulsars detected in the survey}\label{sec:app}
In all, 726 previously-known pulsars were re-detected in the mid-latitude survey data. Their details are listed in the online supporting material\footnote{http://assets.slate.wvu.edu/resources/261/1346789669.pdf}, which also includes modifications to the published period and DM values, in some cases. Where we have modified the pulse period, the published value was in fact a multiple of the true period. In the case of PSR~J0905$-$4536, the published DM value is 116.8~$\rm{cm}^{-3}\,\rm{pc}$, but when folding the data, it was clear that the true DM is in fact much higher, 179.7~$\rm{cm}^{-3}\,\rm{pc}$. The reason for this error is unclear.

A further 96 pulsars in the region were too weak to be detected in a blind search at this sensitivity limit, but were detected when the data were folded with the correct parameters. There were, however, 70 pulsars that were not detected despite being relatively bright. Further inspection shows that strong RFI during observations at these positions caused the pulsars to be obscured.

\bsp

\label{lastpage}

\end{document}